\documentclass{article}
\usepackage{amsfonts,amsmath,amssymb,bookmark,epsfig,float,graphicx,here,latexsym,mathtools,setspace}
\usepackage[all]{xy}
\usepackage[usenames]{color}
\usepackage{comment}
\usepackage{subcaption}
\usepackage{tikz}
\usepackage{pgfplots}
\usepackage{adjustbox}
\usepackage{natbib}
\usepackage{pgfplots}
\usepackage{eurosym}
\pgfplotsset{compat=1.18} 
\newtheorem{theorem}{Theorem}[section]

\newtheorem{assumption}[theorem]{Assumption}

\newtheorem{corollary}[theorem]{Corollary}

\newtheorem{definition}[theorem]{Definition}
\newtheorem{example}[theorem]{Example}

\newtheorem{lemma}[theorem]{Lemma}

\newtheorem{remark}[theorem]{Remark}

\usepackage[a4paper, top=3cm, bottom=3.5cm, left=3.5cm, right=3.5cm]{geometry}

\newenvironment{proof}[1][Proof]{\noindent\textbf{#1.} }{\ \rule{0.5em}{0.5em}}

\newcommand{\R}{\mathbb{R}}
\newcommand{\E}{\operatorname{\mathbb{E}}}
\newcommand{\diff}{\mathrm{d}}

\newcommand{\sqb}[1]{\ensuremath{ \left[ #1 \right] }}
\newcommand{\of}[1]{\ensuremath{\left( #1 \right)}}
\newcommand{\cb}[1]{\ensuremath{ \left\{ #1 \right\} }}

\DeclareMathOperator*{\argmin}{arg\,min}

\title{Higher-Order Ambiguity Attitudes\thanks{We are very grateful to Louis Eeckhoudt and seminar participants at Bayes Business School London for their comments and suggestions. 
This research was supported in part by the Netherlands Organization for Scientific Research under grant NWO Vici 2020--2027 (Ayg\"un and Laeven).}}
\author{M\"ucahit Ayg\"un\\
{\footnotesize Dept.~of Quantitative Economics}\\
{\footnotesize University of Amsterdam }\\
{\footnotesize and Tinbergen Institute}\\
{\footnotesize {\tt M.Aygun@uva.nl}}\\[1mm]
\and Roger J.~A. Laeven\thanks{Corresponding author. Mailing Address: PO Box 15867, 1001 NJ Amsterdam, The Netherlands.
Phone: +31 (0) 20 525 4252.}\\
{\footnotesize Dept.~of Quantitative Economics}\\
{\footnotesize University of Amsterdam, EURANDOM}\\
{\footnotesize and CentER}\\
{\footnotesize {\tt R.J.A.Laeven@uva.nl}}\\[1mm]
\and Mitja Stadje\\
{\footnotesize Inst. of Insurance Science and Inst. of Financial Mathematics}\\
{\footnotesize Faculty of Mathematics and Economics}\\
{\footnotesize Ulm University}\\
{\footnotesize {\tt Mitja.Stadje@uni-ulm.de}}}

\date{First version: April 23, 2023\\
This version: \today}

\begin{document}

\maketitle

\begin{abstract}
We introduce a model-free preference under ambiguity, as a primitive trait of behavior, which we apply once as well as repeatedly. 
Its single and double 
application yield simple, easily interpretable definitions of ambiguity aversion and ambiguity prudence.  
We derive their implications within canonical models for decision under risk and ambiguity.
We establish in particular that our new definition of ambiguity prudence is equivalent to a positive third derivative of: (i) the capacity in the Choquet expected utility model (\citealp{S89}), (ii) the dual conjugate of the divergence function under variational divergence preferences (\citealp{MMR06}) and (iii) the ambiguity attitude function in the smooth ambiguity model (\citealp{KMM05}).  
We show that our definition of ambiguity prudent behavior may be naturally linked to an optimal insurance problem under ambiguity.
\end{abstract}


\newpage\section{Introduction}\label{sec:int}

Attitudes toward ambiguity (probabilities unknown; \citealp{Keynes21,Knight21})
nowadays play a central role in decision-making under uncertainty,
almost on equal footing with attitudes toward risk (probabilities given).
Over the past four decades, a variety of decision models have been developed that can accommodate \cite{E61} type phenomena
by forgoing the assignment of subjective probabilities (\citealp{R31}, \citealp{dF31})
as in the subjective expected utility (SEU) model of \cite{S54}.
Canonical examples include the Choquet expected utility (CEU) model of \cite{S89},
the maxmin expected utility (MEU) model of \cite{GS89},
the multiplier preferences (MP) model of \cite{HS01},
the variational preferences (VP) model of \cite{MMR06},
the second-order expected utility (SOEU) model of \cite{N93,N10}
and the smooth ambiguity (SA) model of \cite{KMM05}.\footnote{These models can be unified as models displaying uncertainty averse preferences: complete, transitive, monotone and convex preferences (\citealp{CMMM11}). 
Dual and rank-dependent extensions and generalizations of the MEU, MP and VP models have been axiomatized in \cite{LS23}.}

In this paper, we introduce a model-free ambiguity preference --- a primitive trait of behavior with a simple interpretation --- and analyze its implications within the various models for decision under uncertainty, when applied once as well as repeatedly.
The one-shot application of our model-free preference yields a simple, primitive definition of aversion to ambiguity; repeated applications yield higher-order ambiguity attitudes, starting with ambiguity prudence 
at the third 
order.
We establish in particular that our definition of ambiguity prudence is equivalent to a positive sign of the third derivative of: the capacity under CEU; 
the dual conjugate of the divergence function under variational divergence preferences; 
the ambiguity attitude function under SOEU and SA; and is automatically satisfied under MP.
Furthermore, we demonstrate that ambiguity aversion is equivalent to a positive sign of the second derivative of the capacity under CEU; a concave ambiguity attitude function under SOEU and SA; and is always satisfied under MEU, MP and VP. 
This paper thus links the pivotal elements of theories for decision under uncertainty --- capacities (CEU), ambiguity indexes (VP), ambiguity attitude functions (SA) --- to model-free preferences (rather than taking them as exogenously given). 

Our model-free preference can be understood as follows.
Suppose there are (at least) two states of the world, with unknown probabilities of occurring. 
Imagine a decision-maker who is indifferent between permutations of these states of the world.
For a given stochastic endowment, a state is considered to be good (bad) if it yields a high (low) utility.
Our simple model-free preference then asserts that 
the decision-maker prefers experiencing a utility loss (gain) in a good (bad) state to experiencing an equal-sized utility loss (gain) in a bad (good) state.
The repeated application of our model-free preference, requiring at least three states, asserts the preference for a utility gain to occur in a bad state and an equal-sized utility loss to occur in an intermediate state
rather than experiencing the utility gain in the intermediate state and the utility loss in a good state.
In other words, she prefers to combine good --- an attractive transformation --- with bad --- an unfavorable part of the state space.
Thus, our model-free preference admits a simple `hedging' interpretation.
Note the nested nature of our approach, where the third order transformations are defined by applying the second order transformations twice.

Our analysis involves first-, second- and higher-order derivatives of capacities (i.e., monotone and grounded set functions).
Under CEU, our definition of ambiguity aversion agrees with the classical uncertainty aversion concept of \cite{S89} based on randomization, where it is equivalent to convexity (or better: supermodularity) of the capacity.\footnote{Supermodularity of the capacity can in turn be represented by a positive second order derivative of the capacity. 
See also Corollary~2.23 and Remark~2.24 in \cite{G16} for the connection between the second derivative of the capacity and supermodularity, and for remarks about terminology.}
Contrary to Schmeidler's approach, our construction can also sign the higher-order derivatives of the capacity.
In view of their importance in combinatorial optimization, decision-making and cooperative game theory, uncovering a simple, primitive economic interpretation of the higher-order derivatives of capacities may be viewed as a novel contribution of independent interest.


We show that ambiguity prudence as defined in this paper is naturally connected to an optimal insurance problem with an ambiguous loss.
More specifically, we demonstrate that, in the presence of ambiguity with respect to the loss amount, an ambiguity prudent decision-maker demands more insurance when compared to the situation in which the exact size of the loss, in case it occurs, is known in advance.
That is, an ambiguous loss amount stimulates insurance propensity whenever the decision-maker displays ambiguity prudence.

Our developments on ambiguity attitudes have a parallel under risk.
Indeed, over the past forty years, higher-order risk attitudes have played an increasingly prominent role in decision under risk.
The term prudence (for risk) can be traced back to \cite{K90} who revealed the connection between precautionary savings and a positive sign of the third derivative of the utility function within the expected utility (EU) model.
The signs of the higher-order derivatives of the utility function can however also be given a simple interpretation outside the specific savings problem.
Such a more primitive interpretation was first provided in \cite{MGT80} (at the third order) and developed further by \cite{ES06} under the name of risk apportionment. 
They show, for example, that prudent expected utility maximizers prefer to ``disaggregate the harms'' of a sure reduction in wealth and a zero-mean risk.
\cite{EST09} show that this entails a preference to ``combine good with bad'' instead of facing either everything good or everything bad, where good and bad are defined via stochastic dominance. 
Although presented as a model-free preference, \cite{ELS20} and \cite{vBLvdK23} demonstrate that the \cite{ES06} construction is tailored specifically to the expected utility (EU) model and is not suitable for non-EU models.
\cite{ELS20} and \cite{vBLvdK23} also show how to conduct risk apportionment within \cite{Y87}'s dual theory of choice and \cite{Q82}'s rank-dependent utility model, respectively.
Their dual story retains the generic preference of combining good with bad but departs significantly from the \cite{ES06} approach in its exact implementation, by introducing ``squeezing'' and ``anti-squeezing'' and referring to ``dual moments''.
The simplicity of these approaches makes them particularly amenable to experimental measurements; see e.g., \cite{EW11,EW14}, \cite{DS10,DS14}, \cite{NTvdK14}, \cite{vBLvdK23}.

\cite{B17} extends the \cite{ES06} approach from risk to ambiguity.
Because the \cite{ES06} construction is specifically tailored to the EU model under risk, \cite{B17}'s approach is particularly appealing for the SOEU model for risk and ambiguity, which asserts subjective second-order probabilities over risks and a modified version of SEU.
At the second order, the implications of this construction diverge from those of uncertainty aversion \`a la \cite{S89}. 
That is, at the second order, \cite{B17}'s definition is not equivalent to supermodularity of the capacity under CEU.

Just like there is a well-known link between prudence for risk and decreasing absolute risk aversion (DARA) --- DARA requires prudence as a weaker, necessary (but not sufficient) condition (\citealp[Chapters 1 and 6]{EGS05}) ---,  
similar links may be established between ambiguity prudence as introduced in this paper and decreasing (increasing) absolute (relative) ambiguity aversion.
The latter concept has been analyzed in \cite{CMM22}.
Increasing relative ambiguity aversion is intimately related to a star-shapedness property (\citealp{CCMTW22,LRGZ23,FPS24}), which is gaining traction.

The outline of this paper is as follows.
In Section~\ref{sec:prel}, we provide some preliminaries.
In Section~\ref{sec:modelfree}, we introduce our model-free preference.
Sections~\ref{sec:CEU}--\ref{sec:SA} analyze the implications of the model-free preference within CEU, VP, MEU and SA, and within popular subclasses thereof.
Applications are contained in Section~\ref{sec: Insurance g-div}. 
Section~\ref{sec:con} provides concluding remarks.
All proofs are relegated to the Appendix.

\section{Preliminaries}\label{sec:prel}

\subsection{Setting and Notation}

Consider a finite set $S=\{\omega_1,\ldots,\omega_k\}$ of states of the world with $\sigma$-algebra 
$\Sigma$ that consists of subsets called \textit{events}.
We take $\Sigma$ to be the power set.  
We define a non-empty set $Z$ of \textit{consequences} and denote by $\mathcal{F}$ the set of all the \textit{acts}: finite-valued and $\Sigma$-measurable functions $X:S\rightarrow Z$. 
If $X(\omega_i)=x_i$, with $k=|S|$ the cardinality of $S$, and $x_i\in Z$, we also write 
\begin{equation*}\label{eq:consequences}
	X:=[x_1,\omega_1;\ldots; x_k,\omega_k],
\end{equation*} 
with $x_1\leq \ldots\leq x_k$. 
For a set $E\subset S$ with $\omega_1,\omega_2\notin E$, we let 
$$X(\omega_1,\omega_2):=[\ldots,E;x_1,\omega_1;x_2,\omega_2;\ldots,E^c\setminus\{\omega_1,\omega_2\}];$$ 
i.e., the act $X(\omega_1,\omega_2)$ yields outcome $x_i$ when $\omega_i$ occurs
for $i=1,2$ and arbitrary, smaller and larger outcomes whenever $E$ and $E^c\setminus\{\omega_1,\omega_2\}$ happen.
For $x\in Z$, with the usual slight abuse of notation, we define $x\in\mathcal{F}$ to be the constant act such that $x(s)=x$ for all $s\in S$. 
Suppose the decision-maker's (DM) preferences are given by a binary relation $\succcurlyeq$ on $\mathcal{F}$, where, as usual, $\succ$ stands for strict preference and $\sim$ for indifference. 
Throughout the paper, we assume that $u(\cdot)$ is a utility function, describing the degree of satisfaction 
of the consequence $\cdot$, induced by a numerical representation on the space of constant acts.

We assume that all states $\omega$ are interchangeable in the sense that the DM has no reason \textit{a priori} to consider an outcome in a specific state to be more valuable than in another one, i.e.:
\begin{assumption}\label{ass:sym} 
For all $E\subset S$,
$$[\ldots,E;u_1,\omega_1;u_2,\omega_2;\ldots,E^c\setminus\{\omega_1,\omega_2\}]\sim [\ldots,E;u_2,\omega_1;u_1,\omega_2;\ldots,E^c\setminus\{\omega_1,\omega_2\}].$$ 
\end{assumption}
Since our approach is probability-wise model free, such a symmetry assumption is necessary in order to analyze choices. 
Note that this entails that the DM is indifferent regarding all permutations of $u_1,\ldots, u_k$.

\subsection{Capacities and Their Derivatives}

Our analysis involves capacities and their first- and higher-order derivatives. 

We define set functions on a nonempty finite set $S$ with domain the power set $2^{S}$.
Hence, a set function $\nu$ on $S$ is a mapping $\nu:2^{S}\rightarrow\mathbb{R}$.
A \textit{capacity} is a monotone and grounded set function, that is, a capacity $\nu$ is a set function satisfying $\nu(A)\leq \nu(B)$ if $A\subseteq B$, $A,B\in 2^{S}$, and $\nu(\emptyset)=0$.

Consider a set function $\nu$ on $S$.
The \textit{derivative} of the set function $\nu$ at $A\subseteq S$ w.r.t.\ $\omega\in S$ is defined by
\begin{equation*}
\Delta_{\omega}\nu(A)=\nu(A\cup\{\omega\})-\nu(A\setminus\{\omega\}).
\end{equation*}
Observe that $\Delta_{\omega}\nu(A\cup\{\omega\})=\Delta_{\omega}\nu(A\setminus\{\omega\})$. 
That is, the presence or absence of $\omega$ in the set $A$ has no effect on the resulting derivative.
Higher-order derivatives are defined as follows.
The second-order derivative of $\nu$ at $A\subseteq S$ w.r.t.\ $\omega_{1},\omega_{2}\in S$ is defined by
\begin{align*}
\Delta_{\omega_{1},\omega_{2}}\nu(A)&=\Delta_{\omega_{1}}\left(\Delta_{\omega_{2}}\nu(A)\right)=\Delta_{\omega_{2}}\left(\Delta_{\omega_{1}}\nu(A)\right)\\
&=\nu(A\cup\{\omega_{1},\omega_{2}\})-\nu(A\cup\{\omega_{1}\}\setminus\{\omega_{2}\})\\
&\quad-\nu(A\cup\{\omega_{2}\}\setminus\{\omega_{1}\})+\nu(A\setminus\{\omega_{1},\omega_{2}\}),
\end{align*}
see \citet[Chapter~2]{G16}.
Similarly, the $n$-th order derivative of $\nu$ at $A\subseteq S$ w.r.t.\ $\omega_{1},\ldots,\omega_{n}\in S$ is defined iteratively by
\begin{equation*}
\Delta_{\omega_{1},\ldots,\omega_{n}}\nu(A)=\Delta_{\omega_{1}}\left(\ldots\Delta_{\omega_{n}}\nu(A)\ldots\right).
\end{equation*}
In particular, at the third order,
\begin{align*}
&\Delta_{\omega_{1},\omega_{2},\omega_{3}}\nu(A)\\
&=\nu(A\cup\{\omega_{1},\omega_{2},\omega_{3}\})\\
&\quad-\nu(A\cup\{\omega_{1},\omega_{2}\}\setminus\{\omega_{3}\})-\nu(A\cup\{\omega_{1},\omega_{3}\}\setminus\{\omega_{2}\})-\nu(A\cup\{\omega_{2},\omega_{3}\}\setminus\{\omega_{1}\})\\
&\quad+\nu(A\cup\{\omega_{1}\}\setminus\{\omega_{2},\omega_{3}\})-\nu(A\cup\{\omega_{2}\}\setminus\{\omega_{1},\omega_{3}\})-\nu(A\cup\{\omega_{3}\}\setminus\{\omega_{1},\omega_{2}\})\\
&\quad-\nu(A\setminus\{\omega_{1},\omega_{2},\omega_{3}\}).
\end{align*}

More generally, for any set $B\subseteq S$, the derivative of $\nu$ at $A\subseteq S$ w.r.t.\ $B$ can be found as
\[\Delta_B\nu(A)=\sum_{C\subseteq B}(-1)^{|B\setminus C|}\nu((A\setminus B)\cup C).\]


The following lemma highlights basic properties of the derivatives of capacities; it is based on Ch.\ 2 in \cite{G16}.
\begin{lemma}
For a capacity $\nu:2^S\to \R$, the following statements hold:
    \begin{itemize}
         \item [(i)] The first derivative $\Delta_{\omega}\nu(A)$ is non-negative for any set $A\subseteq S$ and $\omega\in S$.
         \item [(ii)] The second derivative $\Delta_{\omega_{1},\omega_{2}}\nu(A)$ is non-negative (non-positive) for any set $A\subseteq S$ and $\omega_1,\omega_2 \in S$ if and only if $\nu$ is supermodular (submodular), i.e., $\nu(A\cup B)+\nu(A\cap B)\geq (\leq)\, \nu(A)+\nu(B)$ for every $A,B\subseteq S$.
         \item [(iii)] For every $A,B\subseteq S$ with $|B|\leq n$, $\Delta_B \nu(A)\geq 0$ if and only if $\nu$ is $n$-monotone, i.e., \[\nu\of{\bigcup_{i=1}^n A_i}\geq \sum_{I\subseteq \{1,2,\ldots,n\}, I\neq \emptyset}(-1)^{|I|+1}\nu\of{\bigcap_{i\in I}A_i},\]
         for any family of sets $A_1,A_2,\ldots A_n\subseteq S$.
    \end{itemize}
\end{lemma}
 
Consider an act $X$ that takes the values $u_{1},\ldots,u_{n}$
with $0\leq u_{1}<\cdots< u_{n}$.
The \emph{Choquet integral} of $X$ w.r.t.\ the capacity $\nu$ and utility function $u$ is given by
\begin{equation}\label{eq:ChoquetIntegral}
V(X):=\sum_{i=1}^{n}\nu(A_{i})(u_{i}-u_{i-1}),
\end{equation}
with $0=u_{0}\leq u_{1}<\cdots< u_{n}$ and $A_{i}=\{\omega\in S|u(X(\omega))\geq u_{i}\}$, $i=1,\ldots,n$.
Here, $A_{1}=S$.


\section{Model-Free Preferences}\label{sec:modelfree}
In this section, we introduce a simple model-free preference, as a primitive trait of behavior.
Let us consider three acts, $X_I^{(2)}$, $X_A^{(2)}$ and $X_B^{(2)}$, defined as follows:
\begin{align}
u(X_{I}^{(2)}(\omega_{1},\omega_{2}))&:=[.,E ; u_1,\omega_1 ; u_2,\omega_2 ; ..,E^{c}\setminus\{\omega_{1},\omega_{2}\}],\label{eq:X_I^2}\\
u(X_{A}^{(2)}(\omega_{1},\omega_{2}))&:=[.,E ; u_1-\bar{u},\omega_1 ; u_2+\bar{u},\omega_2 ; ..,E^{c}\setminus\{\omega_{1},\omega_{2}\}],\label{eq:X_A^2}\\
u(X_{B}^{(2)}(\omega_{1},\omega_{2}))&:=[.,E ; u_1+\bar{u},\omega_1 ; u_2-\bar{u},\omega_2 ; ..,E^{c}\setminus\{\omega_{1},\omega_{2}\}],\label{eq:X_B^2}
\end{align}
with $u_{1}<u_{2}$ and $\bar{u}>0$.
We assume throughout that $\bar{u}$ is such that the ranking of the utility outcomes is preserved.

\begin{definition}\label{def:ambavr}
We say that the DM is ambiguity averse if 
\begin{equation*}
X_{B}^{(2)}(\omega_{1},\omega_{2})\succeq X_{A}^{(2)}(\omega_{1},\omega_{2}),\label{eq:ambavr}    
\end{equation*}
for any $E\subset S$ with $\omega_{1},\omega_{2}\notin E$, $u_{1}<u_{2}$ and $\bar{u}>0$.
\end{definition}
Our preference at the second order may be given a hedging interpretation.
It asserts that an ambiguity averse DM prefers to give up $\bar{u}$ utility units in some state ($\omega_2$) and gain $\bar{u}$ utility units in a worse state ($\omega_1$)
rather than the opposite --- gaining $\bar{u}$ in state $\omega_2$ and losing $\bar{u}$ in state $\omega_1$.

Next, let us consider the three acts $X_I^{(3)}$, $X_A^{(3)}$ and $X_B^{(3)}$ defined by:
\begin{align}
u(X_{I}^{(3)}(\omega_{1},\omega_{2},\omega_{3}))&:=[.,E ; u_1,\omega_1 ; u_2,\omega_2 ; u_3,\omega_3 ; ..,E^{c}\setminus\{\omega_{1},\omega_{2},\omega_{3}\}],\label{eq:X_I^3}\\
u(X_{A}^{(3)}(\omega_{1},\omega_{2},\omega_{3}))&:=[.,E ; u_1-\bar{u},\omega_1 ; u_2+2\bar{u},\omega_2 ; u_3-\bar{u},\omega_3 ; ..,E^{c}\setminus\{\omega_{1},\omega_{2},\omega_{3}\}],\label{eq:X_A^3}\\
u(X_{B}^{(3)}(\omega_{1},\omega_{2},\omega_{3}))&:=[.,E ; u_1+\bar{u},\omega_1 ; u_2-2\bar{u},\omega_2 ; u_3+\bar{u},\omega_3 ; ..,E^{c}\setminus\{\omega_{1},\omega_{2},\omega_{3}\}],\label{eq:X_B^3}
\end{align}
with $u_{1}<u_{2}<u_{3}$, $\bar{u}>0$ and $u_2-u_1 = u_3-u_2=:\Delta u>0$.
We assume again throughout that $\bar{u}$ is such that the ranking of the utility outcomes is preserved.
Note that $X_{A}^{(3)}(\omega_{1},\omega_{2},\omega_{3})$ and $X_{B}^{(3)}(\omega_{1},\omega_{2},\omega_{3})$ are obtained from $X_{I}^{(3)}(\omega_{1},\omega_{2},\omega_{3})$ by conducting the second-order transformation above twice.
More specifically, to obtain $X_{B}^{(3)}(\omega_{1},\omega_{2},\omega_{3})$ we conduct the `good' second-order transformation in the two low-utility states and conduct the `bad' second-order transformation in the two high-utility states.
The opposite applies to $X_{A}^{(3)}(\omega_{1},\omega_{2},\omega_{3})$.

\begin{definition}\label{def:ambprud}
We say that the DM is ambiguity prudent if 
\begin{equation*}
X_{B}^{(3)}(\omega_{1},\omega_{2},\omega_{3})\succeq X_{A}^{(3)}(\omega_{1},\omega_{2},\omega_{3}),\label{eq:ambprud}    
\end{equation*}
for any $E\subset S$ with $\omega_{1},\omega_{2},\omega_{3}\notin E$, $u_{1}<u_{2}<u_{3}$, $\bar{u}>0$ and $\Delta u>0$.
\end{definition}

In view of its simplicity, our behavioral definition of ambiguity prudence admits a straightforward translation into urn examples. 
For instance, suppose we consider an urn containing 90 balls. 
The balls are black, red or yellow, in unknown proportions. 
The balls are well mixed.
Suppose that a ball is drawn from the urn.
The 
payoff is contingent on the color of the drawn ball.
We then have that an ambiguity prudent DM prefers the utility profile 
\begin{table}[H]
\begin{center}
\begin{tabular}{c|ccc}
\hline  
\textit{Color drawn} & Black & Red & Yellow\\
\textit{Utility} & 60 & 60 & 240\\
\hline 
\end{tabular}
\label{tab:B}
\end{center}
\end{table}
\vskip -0.4cm
\noindent to the utility profile 
\begin{table}[H]
\begin{center}
\begin{tabular}{c|ccc}
\hline  
\textit{Color drawn} & Black & Red & Yellow\\
\textit{Utility} & 0 & 180 & 180\\
\hline 
\end{tabular}
\label{tab:A}
\end{center}
\end{table}
\noindent The preferred utility profile is attractive because it 
yields the highest utility 
in the worst state.
It speaks to a DM’s pessimism.
Besides, it also has the highest possible utility. 
The price to pay is a reduction of the utility
in the middle/intermediate state.


\section{Choquet Expected Utility}\label{sec:CEU}

The Choquet expected utility model (CEU) developed in \cite{S89} postulates that an act $X$ taking the monetary values $x_{1}<\cdots< x_{n}$ is evaluated according to 
$$U(X)\equiv V(u(X))=\sum_{i=1}^{n}\nu(A_{i})(u(x_i)-u(x_{i-1})),$$
where $u:\mathbb{R}^{+}_{0}\rightarrow\mathbb{R}$, with $u(x_{0})=0$ for $x_{0}\leq x_{1}$, is a utility function of wealth 
and $\nu:2^{S}\rightarrow \mathbb{R}$ is a capacity.
In the sequel, we will assume a capacity to be normalized such that $\nu(S)=1$.

\subsection{Ambiguity Aversion}

We start by demonstrating that our definition of ambiguity aversion has the same implications as the classical ``uncertainty aversion'' of \cite{S89} within the CEU model.\footnote{This is not the case for the approach of \cite{B17}. 
His definition of ambiguity aversion is linked to ``binary superadditivity'': for all $E\subseteq S$, $\nu(E)+\nu(E^{c})\leq 1$.} 
At the second order, diversification/hedging (our ambiguity aversion) and randomization (uncertainty aversion) agree, from the third order onwards, the two approaches diverge.

Formally, \cite{S89} introduces the notion of uncertainty aversion through a binary preference relation on any three acts $X_1,X_2,X_3$, and any $\alpha\in[0,1]$: if $X_1\succcurlyeq X_3$ and $X_2\succcurlyeq X_3$, then $\alpha X_1+(1-\alpha)X_2\succcurlyeq X_3$. 
Equivalently, one may state: if $X_1\succcurlyeq X_2$, then $\alpha X_1+(1-\alpha)X_2\succcurlyeq X_2$. 
The definition of strict uncertainty aversion asserts that the conclusion is a strict preference $\succ$.\footnote{See, e.g., \cite{C91} for further details on uncertainty aversion and its link to capacities.} 
An uncertainty averse DM prefers to randomize acts, trading off ambiguity for chance.

Recall 
Definition~\ref{def:ambavr}.\footnote{We note that this is a special case of Axiom~A6 in \cite{LS23}.
\cite{LS23} consider subjective mixtures of random variables via utility profiles. 
Indeed, the utility profile of a subjective mixture of two random variables equals the convex combination of the utility profiles of the random variables themselves. 
More specifically, see Eqn.~(4.2) in \cite{LS23}.}
We state the following theorems:

\begin{theorem}\label{th:CEU2nd-i}
If a CEU DM is ambiguity averse, 
then the second derivative of the capacity $\nu$ is non-negative.
\end{theorem}

\begin{theorem}\label{th:CEU2nd-ii}
Grant Assumption~\ref{ass:sym}. 
A CEU DM 
is ambiguity averse provided the second derivative of the capacity $\nu$ is non-negative.
\end{theorem}

\subsubsection{Ambiguous insurance}
A non-negative second derivative of $\nu$ may be given a simple interpretation: it means that eliminating ambiguity becomes more valuable the closer an outcome is to a certain outcome. 
To illustrate, suppose we consider an urn containing $100$ balls of 10 different, enumerated colors.
The number of balls per color is, however, unknown.
(Alternatively, one may assume that the DM has partial information about the number of balls per color --- for instance, each color having a minimal number of $r$ balls --- as long as this information is `symmetric' with respect to each color.)

Suppose a ball is drawn from the urn and the DM loses $\ell=$ \euro\ $2\mathord{,}000$ if a ball with one of the first two colors is drawn. 
In the sequel, these colors shall be referred to as the ``losing colors''. 
They may be regarded as representing two unfavorable events a DM wants to purchase insurance for.
Denote by $X$ the act described above.
Furthermore, let $c$ denote the certainty equivalent of the act $X$.
That is, the DM is indifferent between the act $X$ and paying (i.e., losing) the monetary amount $|c|$ with certainty.

Now imagine the following act: a ball is drawn from the same urn; the DM loses \euro\ $2\mathord{,}000$ if a ball with (only) the first color (rather than the first two colors) is drawn, and loses $|c/2|$, i.e., half the original certainty equivalent in all states of the world.
We denote by $Y$ the act just described.
In insurance terms, paying $|c|$ corresponds to ``full insurance'', bearing $X$ is ``no insurance'', and $Y$ is ``ambiguous insurance''.
Ambiguous insurance is the counterpart of probabilistic insurance; see \cite{KT79,WTT97} for a detailed treatment of probabilistic insurance.\footnote{Under risk, probabilistic insurance is disliked by a DM who complies with \cite{Y87}'s dual theory or \cite{Q82}'s rank-dependent utility model and displays probability-driven risk aversion \citep{CKS87,EL22}. }
In what follows, we assume an affine utility function $u(x)=x+M$, $M\in\mathbb{R}$, so that the dispreference of ambiguous insurance is explained only by the shape of the capacity $\nu$.
Indeed, an ambiguity averse CEU DM, with a convex capacity, will prefer $c$ hence $X$ to $Y$.
Thus, we can find $m>0$ such that $c-m\sim Y$.
We refer to Figure~\ref{fig:amb ins 1} for a graphical representation of $c$, $X$ and $Y$.
\begin{figure}[t]
   \centering
\resizebox{0.85\textwidth}{!}{
    \begin{subfigure}[b]{0.2\textwidth}
        \raisebox{7.5ex}{\begin{tikzpicture}[scale=1, decoration={ticks, segment length=2pt, amplitude=1.8pt}]
		\draw[- >, thick] (0,0) -- (0.9,0) -- (1.5,0) node[right] {$c$};				
  		\draw[mark options={fill=white}]
      plot[mark=*] (0,0);
	\end{tikzpicture}}
    \end{subfigure}
    \hspace{1em}
        \begin{subfigure}[b]{0.05\textwidth}
        \begin{tikzpicture}
        \raisebox{8.5ex}{$\sim$}
    \end{tikzpicture}
\end{subfigure}
\hspace{1em}
    \begin{subfigure}[b]{0.2\textwidth}
	\raisebox{9ex}{\begin{tikzpicture}[scale=1, decoration={ticks, segment length=2pt, amplitude=1.8pt}, baseline ]		
		\draw[- >, thick] (0,0) -- (0.75,0.375) node[rotate=25,above] {\scriptsize $S \setminus \{\omega_1,\omega_2$\}} -- (1.5,0.75) node[right] {$0$};  		
  		\draw[- >, thick] (0,0) -- (0.75,-0.375) node[rotate=-25,below] {\scriptsize $\{\omega_1,\omega_2\}$} --(1.5,-0.75) node[right] {$-\ell$};
  		\draw[mark options={fill=white}]
      plot[mark=*] (0,0);

	\end{tikzpicture}}
    \end{subfigure}
    \hspace{1em}
        \begin{subfigure}[b]{0.05\textwidth}
        \begin{tikzpicture}
        \raisebox{8.5ex}{$\succ$}
    \end{tikzpicture}
\end{subfigure}
\hspace{1em}
     \begin{subfigure}[b]{0.2\textwidth}
	\raisebox{9ex}{\begin{tikzpicture}[scale=1, decoration={ticks, segment length=2pt, amplitude=1.8pt}, baseline ]
		\draw[- >, thick] (0,0) -- (0.75,0.375) node[rotate=25,above] {\scriptsize $S \setminus \{\omega_1,\omega_2$\}} -- (1.5,0.75) node[right] {$c/2$};  		
  		\draw[- > , thick] (0,0) -- (0.75,-0.375) - > (1.5,-0.15) node[rotate = 25,midway,above] {\scriptsize $\omega_2$} node[right] {$c/2$};
        \draw[- > , thick] (0,0) -- (0.75,-0.375) - >(1.5,-0.75) node[rotate = -25,midway,below] {\scriptsize $\omega_1$} node[right] {$c/2-\ell$};
  	\draw[mark options={fill=white}]
      plot[mark=*] (0,0);
 \end{tikzpicture}}
 \end{subfigure}}
\caption{Ambiguous insurance\newline
{\small\textit{Notes}: This figure illustrates the certainty equivalent $c$ (full insurance) and the acts $X$ (no insurance) and $Y$ (ambiguous insurance), where $\ell =$ \euro\  $2\mathord{,}000$.}}
\label{fig:amb ins 1}
\end{figure}

More generally, consider the acts $c(w_0,K)$, $X(w_0,K)$, $Y(w_0,K)$ that yield a large loss of $K\gg\ $\euro\ $2\mathord{,}000$ if one of the $w_0$ \textit{highly} unfavorable colors is drawn, and yield $c,X,Y$ otherwise. 
An ambiguity averse CEU DM will prefer $c(w_0,K)$ hence $X(w_0,K)$ to $Y(w_0,K)$, i.e., there exists $m>0$ such that indifference occurs between the act that pays off $-K$ if one of the $w_0$ highly unfavorable colors is drawn and $c(w_0,K)-m$ otherwise on the one hand and the act $Y(w_{0},K)$ on the other hand.
We refer to Figure~\ref{fig:amb ins 2} for a graphical representation of $c(w_0,K)$, $X(w_0,K)$ and $Y(w_0,K)$.
\begin{figure}[b]
   \centering
 \hspace{1em}
   \resizebox{0.85\textwidth}{!}{
    \begin{subfigure}[b]{0.225\textwidth}
        \raisebox{1ex}{\begin{tikzpicture}[scale=1, decoration={ticks, segment length=2pt, amplitude=1.8pt}]

		\draw[- >, thick] (0,0) -- (2,0) node[,midway,above] {\scriptsize $S\setminus E$} node[right] {$c$};
            \draw[- >, thick] (0,0)  -- (2,-1) node[rotate = -25,midway,below] {\scriptsize $E$}  node[right] {$-K$};
				
  		\draw[mark options={fill=white}]
      plot[mark=*] (0,0);

	\end{tikzpicture}}
    \end{subfigure}
    \hspace{1em}
        \begin{subfigure}[b]{0.05\textwidth}
        \begin{tikzpicture}
        \raisebox{8.5ex}{$\sim$}
    \end{tikzpicture}
\end{subfigure}
\hspace{1em}
    \begin{subfigure}[b]{0.225\textwidth}
	\raisebox{9ex}{\begin{tikzpicture}[scale=1, decoration={ticks, segment length=2pt, amplitude=1.8pt}, baseline ]		
		\draw[- >, thick] (0,0)  -- (2,1) node[rotate=25, midway,above] {\scriptsize $S \setminus (E\cup \{\omega_1,\omega_2$\})} node[right] {$0$};  
          \draw[- >, thick] (0,0) -- (0.8,-0.1) -- (2,-0.25)  node[rotate = -5,midway,above]{\scriptsize $\{\omega_1,\omega_2\}$} node[right] {$-\ell$};
  		\draw[- >, thick] (0,0) -- node[rotate=-25, midway,below]{\scriptsize $E$}(2,-1) node[right] {$-K$};
  		\draw[mark options={fill=white}]
      plot[mark=*] (0,0);

	\end{tikzpicture}}
    \end{subfigure}
    \hspace{1em}
        \begin{subfigure}[b]{0.05\textwidth}
        \begin{tikzpicture}
        \raisebox{8.5ex}{$\succ$}
    \end{tikzpicture}
\end{subfigure}
\hspace{1em}
     \begin{subfigure}[b]{0.225\textwidth}
	\raisebox{9ex}{\begin{tikzpicture}[scale=1, decoration={ticks, segment length=2pt, amplitude=1.8pt}, baseline ]
		\draw[- >, thick] (0,0)  -- (2,1) node[rotate=25, midway,above] {\scriptsize $S \setminus (E\cup \{\omega_1,\omega_2$\})} node[right] {$c/2$};  		
  		\draw[- > , thick] (0,0) -- (1.2,-0.15) - > (2,0.25) node[rotate = 25,midway,above] {\scriptsize $\omega_2$} node[right] {$c/2$};
        \draw[- > , thick] (0,0) -- (1.2,-0.15) - >(2,-0.25) node[rotate = -25,midway,below] {\scriptsize $\omega_1$} node[right] {$c/2-\ell$};
          \draw[- >, thick] (0,0) -- node[rotate=-25, midway,below]{\scriptsize $E$}(2,-1) node[right] {$-K$};
  	\draw[mark options={fill=white}]
      plot[mark=*] (0,0);
 \end{tikzpicture}}
   \end{subfigure}
}
\hspace{1em}
\caption{Ambiguous insurance with a large loss\newline
{\small\textit{Notes}: This figure illustrates the generalized acts $c(w_{0},K)$, $X(w_{0},K)$ and $Y(w_{0},K)$, where $\ell =$ \euro\  $2\mathord{,}000$ and $K\gg \ell$.}}
\label{fig:amb ins 2}
\end{figure}

We can translate this into the following ramifications.
Let $\bar{\nu}(A)=1-\nu(S\setminus A)$ for all $A\subseteq S$.\footnote{We have 
\begin{align}
V(X)&=\sum_{i=1}^{n}\nu(A_{i})(u_{i}-u_{i-1})
=-\sum_{i=1}^{n}u_{i}\left(\nu(A_{i+1})-\nu(A_{i})\right)\nonumber\\
&=\sum_{i=1}^{n}u_{i}\left(\bar{\nu}(S\setminus A_{i+1})-\bar{\nu}(S\setminus A_{i})\right),
\end{align}
where $0=u_{0}\leq u_{1}<\cdots< u_{n}$ and $A_{i}=\{\omega\in S|u(X(\omega))\geq u_{i}\}$, $i=1,\ldots,n$, hence $\nu(A_{n+1})=\nu(\emptyset)=0$.
}
Then, with $M=\ell$,
\begin{align*}
U(c)=c+\ell&=\left(1-\bar{\nu}(\{\omega_1,\omega_2\})\right)\ell\\
&\geq \left(1-\bar{\nu}(\omega_1)-\frac{\bar{\nu}(\{\omega_1,\omega_2\})}{2}\right)\ell\\
&=\frac{c}{2}+\left(1-\bar{\nu}(\omega_1)\right)\ell=U(Y),
\end{align*}
where $\ell=2\mathord{,}000$.
Here, we used the property that 
\[
\bar{\nu}(\omega_1)+\frac{\bar{\nu}(\{\omega_1,\omega_2\})}{2}\geq \bar{\nu}(\{\omega_1,\omega_2\},
\]
which, by Assumption~\ref{ass:sym}, can be re-written as
\[
\bar{\nu}(\omega_1)+\bar{\nu}(\omega_2)\geq \bar{\nu}(\{\omega_1,\omega_2\}.
\]
Furthermore, for the generalized acts, we have
\[
\bar{\nu}(\{\omega_1,E\})+\bar{\nu}(\{\omega_2,E\})\geq \bar{\nu}(\{\omega_1,\omega_2,E\},
\]
where $E$ denotes the event in which the large loss $K$ occurs, 
corresponding to 
the concavity of $\bar{\nu}$ (hence the convexity of $\nu$) by the arbitrariness of $E$.
(Obviously, we could also have chosen another amount than \euro\ $2\mathord{,}000$ to test for ambiguity aversion.)

\subsection{Ambiguity Prudence}

Next, we derive the implications of ambiguity prudence within the CEU model.
The following results reveal that ambiguity prudence is equivalent to a positive sign of the third derivative of the capacity in the CEU model.

Recall 
Definition~\ref{def:ambprud}.
We state the following theorems:

\begin{theorem}\label{th:CEU3rd-i}
If a CEU DM is ambiguity prudent, 
then the third derivative of the capacity $\nu$ is non-negative.
\end{theorem}

\begin{theorem}\label{th:CEU3rd-ii}
Grant Assumption~\ref{ass:sym}. 
A CEU DM 
is ambiguity prudent provided the third derivative of the capacity $\nu$ is non-negative.
\end{theorem}

\begin{remark}
Quite naturally, by requiring Assumption~\ref{ass:sym}, the hypotheses of Theorems~\ref{th:CEU2nd-ii} and~\ref{th:CEU3rd-ii} (and similarly throughout the next sections) require indifference for specific acts that only differ by a permutation of the $\omega$'s and as such rules out situations in which the specific ordering of the $\omega$'s plays a role. 
\end{remark}

\subsubsection{Ambiguous insurance (continued)}
In light of the above interpretation of a non-negative second derivative, a non-negative third derivative of the capacity means that the concavity of the capacity in the states that get insured becomes more pronounced as the multiplicity $w_{0}$ of the uninsured, highly unfavorable states decreases. 
In other words, the value of $m$ that achieves indifference between the act that pays off $-K$ if one of the $w_0$ highly unfavorable colors is drawn and $c(w_0,K)-m$ otherwise on the one hand and the act $Y(w_{0},K)$ on the other hand is decreasing in $w_0$.

\subsection{Neo-Additive CEU}

The neo-additive CEU (NCEU) model has been introduced by \cite{CEG07}. 
It is a special case of the CEU model, in which the capacity $\nu$ is an affine transformation of a probability measure, i.e., given a subjective probability measure $P$, 
\begin{equation*}
\nu(E)=(a-b)/2+(1-a)P(E),
\end{equation*}
for all events $E$ such that $0<P(E)<1$, with $a\in [0,1]$, $b\in [-a,a]$, and $\nu(S)=1-\nu(\emptyset)=1$.

In this model, the probability measure $P$ represents the subjective belief of the DM and $(1-a)$ measures her degree of confidence for this belief. 
In other words, $a$ captures the DM's insensitivity to likelihood. 
The special case of $a\equiv 0$ corresponds to full confidence and subjective expected utility; see also \cite{M11}.
The parameter $b$ is associated with a form of pessimism.
When $b$ is large (small) a small (large) weight will be assigned to the best lottery in expected utility terms.
That is, $(a-b)/2$ measures a level of optimism, with $a=b$ corresponding to pure pessimism and $a=-b$ corresponding to pure optimism.

\begin{remark}\label{rem:NeoCEU-sym}
Grant Assumption~\ref{ass:sym}. 
Then, $P$ is the uniform probability measure.
Indeed, for $u(X_{I}^{(2)}(\omega_{1},\omega_{2}))=[.,E ; u_1,\omega_1 ; u_2,\omega_2 ; ..,E^{c}\setminus\{\omega_{1},\omega_{2}\}]$, choose the set $E$ such that $E\neq \emptyset$ and $E\cup \{\omega_1,\omega_2\}\neq S$. 
Then, using \eqref{eq:ChoquetIntegral}, we have
\begin{align*}
    0&=V(X_I^{(2)}(\omega_1,\omega_2))-V(X_I^{(2)}(\omega_2,\omega_1))
    \\&=(\nu(E^c\setminus \{\omega_1\})-\nu(E^c\setminus \{\omega_2\}))(u_2-u_1)
    \\&= \of{\frac{a-b}{2}+(1-a)P(E^c\setminus \{\omega_1\})-\frac{a-b}{2}-(1-a)P(E^c\setminus \{\omega_2\})}(u_2-u_1).
\end{align*}
Therefore, we obtain $P(E^c\setminus \{\omega_1\})=P(E^c\setminus \{\omega_2\})$. 
Since $\omega_1,\omega_2 \notin E$, using the additivity of probability measures, we find $P(\omega_1)=P(\omega_2)$.
\end{remark}

\subsubsection{Ambiguity Aversion}

We state the following theorem:
\begin{theorem}\label{thm:ambavr-NCEU}
Grant Assumption~\ref{ass:sym}. 
The NCEU DM is ambiguity averse if and only if $a=b$.
\end{theorem}

\begin{figure}[t]
    \centering
    \begin{subfigure}{0.45\textwidth}
    \centering
       \begin{tikzpicture}
            \draw (0,0) rectangle (4,4);
            \draw[blue] (0,0) -- (4,3.2);
            \draw[blue] (4,3.2) circle(0.03cm);
            \fill[white] (4,3.2) circle(0.03cm);
            \filldraw[blue] (4,4) circle(0.03cm);
            \filldraw[black] (-0.2,-0.4) circle (0.0000001pt) node[anchor=south]{0};
            \filldraw[black] (2,-0.6) circle (0.0000001pt) node[anchor=south]{\small $P(E)$};
            \filldraw[black] (4.2,-0.4) circle (0.0000001pt) node[anchor=south]{1};
            \filldraw[black] (-0.4,1.8) circle (0.0000001pt) node[anchor=south]{\small $\nu(E)$\ \ };
            \filldraw[black] (-0.3,4.2) circle (0.0000001pt) node[anchor=west]{1};            
        \end{tikzpicture} 
        \caption{$a=b=0.2$}
    \end{subfigure}
    \begin{subfigure}{0.45\textwidth}
    \centering
       \begin{tikzpicture}
            \draw (0,0) rectangle (4,4);
            \draw[blue] (0,0) -- (4,2);
            \draw[blue] (4,2) circle(0.03cm);
            \fill[white] (4,2) circle(0.03cm);
            \filldraw[blue] (4,4) circle(0.03cm);
            \filldraw[black] (-0.2,-0.4) circle (0.0000001pt) node[anchor=south]{0};
            \filldraw[black] (2,-0.6) circle (0.0000001pt) node[anchor=south]{\small $P(E)$};
            \filldraw[black] (4.2,-0.4) circle (0.0000001pt) node[anchor=south]{1};
            \filldraw[black] (-0.4,1.8) circle (0.0000001pt) node[anchor=south]{\small $\nu(E)$\ \ };
            \filldraw[black] (-0.3,4.2) circle (0.0000001pt) node[anchor=west]{1};  
        \end{tikzpicture} 
        \caption{$a=b=0.5$}
    \end{subfigure}
    \caption{Ambiguity averse neo-additive capacities. \newline
{\small \textit{Notes}: This figure plots two ambiguity averse neo-additive capacities. 
Theorems~\ref{th:CEU2nd-i} and~\ref{th:CEU2nd-ii} yield that a CEU DM is ambiguity averse if and only if the capacity is supermodular. 
To ensure the supermodularity of the neo-additive capacity, the capacity must start in the origin. 
This requires the equality of the parameters $a$ and $b$, as stated in Theorem~\ref{thm:ambavr-NCEU}.}}
\label{fig: Ambiguity averse neo}
\end{figure}

Theorem~\ref{thm:ambavr-NCEU} demonstrates that within NCEU ambiguity aversion is equivalent to $a=b$. 
Therefore, for an NCEU DM, ambiguity aversion is equivalent to pure pessimism.
For ambiguity aversion, from Theorems~\ref{th:CEU2nd-i} and~\ref{th:CEU2nd-ii}, supermodularity of the neo-additive capacity is required. 
Theorem~\ref{thm:ambavr-NCEU} shows that this is only possible when $a=b$. 
The supermodularity can be observed in Figure~\ref{fig: Ambiguity averse neo} for some specific values of the parameters $a$ and $b$.\footnote{In \cite{B17}, his specific form of ambiguity aversion is characterized within NCEU by $b\geq 0$. 
Theorem~\ref{thm:ambavr-NCEU} shows that our definition of ambiguity aversion is equivalent to $a=b$, which already inherits $b\geq 0$. 
Therefore, the implications of the two definitions of ambiguity aversion clearly diverge for the NCEU model.}

\subsubsection{Ambiguity Prudence}
Next, we analyze the implications of ambiguity prudence.
We state the following theorem:
\begin{theorem}\label{thm:ambprud-NCEU}
Grant Assumption~\ref{ass:sym}. 
The NCEU DM is ambiguity prudent.
\end{theorem}

Theorem~\ref{thm:ambprud-NCEU} reveals that an NCEU DM is always ambiguity prudent.
From Theorems~\ref{thm:ambavr-NCEU} and~\ref{thm:ambprud-NCEU}, we conclude that the NCEU DM is always ambiguity prudent but she is ambiguity averse only when she exhibits pure pessimism. 
As we observe from Figure~\ref{fig: Ambiguity prudence neo}, provided the neo-additive capacity is discontinuous at the origin and at $1$, it displays an inverse S-shape.\footnote{
Whereas the implications of the two different definitions of ambiguity aversion in this paper and in \cite{B17} diverge under the NCEU model, 
Theorem~6 in \cite{B17} and Theorem~\ref{thm:ambprud-NCEU} above yield the same characterization of the two different definitions of ambiguity prudence for the NCEU model.
}

\begin{figure}[t]
    \centering
    \begin{subfigure}{0.45\textwidth}
    \centering
       \begin{tikzpicture}
            \draw (0,0) rectangle (4,4);
            \draw[blue] (0,0.4) -- (4,3.2);
            \draw[blue] (4,3.2) circle(0.03cm);
            \draw[blue] (0,0.4) circle(0.03cm);
            \fill[white] (4,3.2) circle(0.03cm);
            \fill[white] (0,0.4) circle(0.03cm);
            \filldraw[blue] (4,4) circle(0.03cm);
            \filldraw[blue] (0,0) circle(0.03cm);
            \filldraw[black] (-0.1,-0.5) circle (0.0000001pt) node[anchor=south]{0};
            \filldraw[black] (2,-0.6) circle (0.0000001pt) node[anchor=south]{\small $P(E)$};
            \filldraw[black] (4,-0.5) circle (0.0000001pt) node[anchor=south]{1};
            \filldraw[black] (-0.4,1.8) circle (0.0000001pt) node[anchor=south]{\small $\nu(E)$\ \ };
            \filldraw[black] (-0.35,4) circle (0.0000001pt) node[anchor=west]{1};            
        \end{tikzpicture} 
        \caption{$a=0.3, b=0.1$}
    \end{subfigure}
    \begin{subfigure}{0.45\textwidth}
    \centering
       \begin{tikzpicture}
            \draw (0,0) rectangle (4,4);
            \draw[blue] (0,0.8) -- (4,4);
            \draw[blue] (0,0.8) circle(0.03cm);
            \fill[white] (0,0.8) circle(0.03cm);
            \filldraw[blue] (4,4) circle(0.03cm);
            \filldraw[blue] (0,0) circle(0.03cm);
            \filldraw[black] (-0.1,-0.5) circle (0.0000001pt) node[anchor=south]{0};
            \filldraw[black] (2,-0.6) circle (0.0000001pt) node[anchor=south]{\small $P(E)$};
            \filldraw[black] (4,-0.5) circle (0.0000001pt) node[anchor=south]{1};
            \filldraw[black] (-0.4,1.8) circle (0.0000001pt) node[anchor=south]{\small $\nu(E)$\ \ };
            \filldraw[black] (-0.35,4) circle (0.0000001pt) node[anchor=west]{1};  
        \end{tikzpicture} 
        \caption{$a=0.2, b=-0.2$}
    \end{subfigure}
\caption{Ambiguity prudent neo-additive capacities. \newline
{\small \textit{Notes}: This figure plots three neo-additive capacities with different parameters. 
Theorems~\ref{th:CEU3rd-i} and~\ref{th:CEU3rd-ii} yield that a CEU DM is prudent if and only if the third derivative of the capacity is non-negative. 
From Theorem~\ref{thm:ambprud-NCEU}, we know that this is always satisfied for neo-additive capacities, also when the capacity does not start in the origin and even when the capacity is concave, as in the figure.}}
\label{fig: Ambiguity prudence neo}
\end{figure}

\section{Variational Preferences}\label{sec:Var}
The variational preferences (VP) model developed by \cite{MMR06} admits, for an act $X$ designated in monetary units, a numerical representation of the form
\begin{equation}\label{eq:VP}
	U(X)=\inf_{Q\in \Delta}\{\sum_{i=1}^{k}q_{i}u(x_{i})+c(Q)\}=\min_{Q \in \Delta}\{\sum_{i=1}^{k}q_{i}u(x_{i})+c(Q)\},
\end{equation}
where $\Delta$ is the set of finitely additive probability measures and 
$c:\Delta\rightarrow\mathbb{R}_+$ is a lower-semicontinuous, convex function referred to as the ambiguity index.
Because $c$ is lower-semicontinuous, the infimum in \eqref{eq:VP} is attained.

We state the following theorem:
\begin{theorem}
   \label{thm: indifference-sym}  
A VP DM satisfies Assumption~\ref{ass:sym}
if and only if $c$ is symmetric with respect to any two arguments, i.e., $c(.,p_1,p_2,.)=c(.,p_2,p_1,.)$ for all $p_1,p_2$. 
\end{theorem}

\subsection{Ambiguity Aversion}
In this subsection, we characterize ambiguity aversion within the VP model. 
Recall Definition~\ref{def:ambavr}.
For an act $u(X)=[.,E;y_1,\omega_1; y_2,\omega_2;..,E^{c}\setminus\{\omega_{1},\omega_{2}\}]$ with $y_1\leq y_2$,
\begin{equation}
\label{eq:VPU2}
U(X)=\min_{(.,p_1,p_2,.)\in \Delta}\{.+p_1 y_1+p_2 y_2+.+c(.,p_1,p_2,.)\}.  
\end{equation}
Denote an argmin of \eqref{eq:VPU2} by $$p^*(y_1,y_2)=(.,p^*_1(y_1,y_2),p^*_2(y_1,y_2),.).$$
In the case the argmin is not unique, denote by $p^*$ an arbitrary $p$ that attains the minimum.

The next theorem shows that within the VP model, ambiguity aversion simply corresponds to assigning larger probabilities to unfavorable scenarios than to favorable scenarios. 
We will see below that any DM complying with VP is ambiguity averse without any further requirements.
\begin{theorem}\label{th:VPaa}
Grant Assumption~\ref{ass:sym}.        
A VP DM is ambiguity averse if and only if 
\begin{equation}
\label{eq:statementAAforVP}
p^*_2(y_1,y_2)\leq  p^*_1(y_1,y_2),
\end{equation}
for any argmin $p^*$.
\end{theorem}

\begin{remark}\label{rem:VPaa}
For a VP DM, \eqref{eq:statementAAforVP} holds for any argmin. 
Indeed, assume that \eqref{eq:statementAAforVP} does not hold, i.e., $p_1^{\ast}(y_1,y_2)<p_2^{\ast}(y_1,y_2)$. 
Then, using Theorem~\ref{thm: indifference-sym}, we have
\begin{align*}
     &.+p_2^{\ast}(y_1,y_2)y_1+p_1^{\ast}(y_1,y_2)y_2+.+c(.,p_2^*(y_1,y_2),p_1^*(y_1,y_2),.)
     \\&= .+p_2^{\ast}(y_1,y_2)y_1+p_1^{\ast}(y_1,y_2)y_2+.+c(.,p_1^*(y_1,y_2),p_2^*(y_1,y_2),.)
     \\&<.+p_1^{\ast}(y_1,y_2)y_1+p_2^{\ast}(y_1,y_2)y_2+.+c(.,p_1^*(y_1,y_2),p_2^*(y_1,y_2),.),
\end{align*}
which contradicts the minimality of $(p_1^{\ast}(y_1,y_2),p_2^{\ast}(y_1,y_2))$. 
Therefore, a VP DM is always ambiguity averse.
\end{remark}

\subsection{Ambiguity Prudence}
Next, we analyze the implications of ambiguity prudence within the VP model.
Recall Definition~\ref{def:ambprud}.
For
\[u(X)=[.,E;y_1,\omega_1; y_2,\omega_2;y_3,\omega_3;..,E^{c}\setminus\{\omega_{1},\omega_{2},\omega_{3}\}]\] 
with $y_1\leq y_2\leq y_3$,
\begin{equation}
    \label{eq:VPU3}
    U(X)=\min_{(.,p_1,p_2,p_3,.)\in \Delta}\{.+p_1 y_1+p_2 y_2+p_3 y_3+.+c(.,p_1,p_2,p_3,.)\}.
\end{equation} 
Denote an argmin of \eqref{eq:VPU3} by 
$$p^*(y_1,y_2,y_3)=(.,p^*_1(y_1,y_2,y_3),p^*_2(y_1,y_2,y_3),p^*_3(y_1,y_2,y_3),.).$$


We state the following theorems: 
\begin{theorem}
\label{th:VP-prudent-necessary}
Grant Assumption~\ref{ass:sym}.        
A VP DM is ambiguity prudent only if 
we have convexity of the minimizing probabilities, i.e.,
\begin{equation}
p^*_2(y_1,y_2,y_3) \leq  \frac{p^*_1(y_1,y_2,y_3)+p^*_3(y_1,y_2,y_3)}{2},\label{eq:con}
\end{equation}
for any argmin $p^*$ whenever $y_2-y_1 > y_3-y_2$ with $y_1\leq y_2\leq y_3$.
\end{theorem}

\begin{theorem}
\label{th:VP-prudent}
Grant Assumption~\ref{ass:sym}.        
A VP DM is ambiguity prudent if 
\begin{equation}
    \label{eq:statement for VP}
p^*_2(y_1,y_2,y_3)\leq  \frac{p^*_1(y_1,y_2,y_3)+p^*_3(y_1,y_2,y_3)}{2}
\end{equation}
for any argmin $p^*$.
\end{theorem}

\begin{remark}
Actually, condition~\eqref{eq:statement for VP} entails that the DM prefers $X_B^{(3)}$ to $X_A^{(3)}$ even without necessarily having $u_2=\frac{u_1+u_3}{2}$ in \eqref{eq:X_I^3}--\eqref{eq:X_B^3} hence in Definition~\ref{def:ambprud}.
If the restriction $u_2=\frac{u_1+u_3}{2}$ is omitted in the definition of ambiguity prudence, then condition~\eqref{eq:statement for VP} is not only sufficient but also necessary for ambiguity prudence.
\end{remark}

These theorems show that, within VP, ambiguity prudence means that the average of the probabilities assigned to the relative worst and relative best states is larger than the probability assigned to the intermediate state. 
Given $p_1,p_2, \ldots, p_k$, denote by $p_{(1)}$ the largest $p_i$, by $p_{(2)}$ the second largest, and by $p_{(k)}$ the smallest $p_i$, i.e., $$p_{(k)}\leq \ldots \leq p_{(2)}\leq p_{(1)}.$$ 
We note that condition~\eqref{eq:statement for VP} is actually equivalent to the function 
$$i\mapsto p^*_{(i)}, $$
mapping from $\{1,\ldots,k\}$ to $[0,1]$, being convex. 
Hence, while ambiguity aversion is equivalent to the function $p_{(i)}$ being decreasing, ambiguity prudence corresponds to $p_{(i)}$ being convex; see Figure~\ref{fig:convexprob} for a graphical representation. 
We are not aware of other works that have investigated such a convexity condition on ordered probability vectors.\footnote{It is interesting to note that the notions of outcome prudence and probability prudence under risk are also connected with a \emph{convexity} condition but w.r.t.\ marginal utility and probability weighting, as represented by the first derivative of the utility and probability weighting functions. 
In an EU setting, this convexity condition translates into a preference for precautionary savings \citep{K90}, whereas under the dual theory \citep{Y87} it is connected to a self-protection problem \citep{ELS20}.}

\begin{figure}
\centering
\begin{tikzpicture}

    \begin{axis}[
        xlabel={$i$},
        ylabel={$p^*_{(i)}$},
        xmin=0,xmax=10,
        ymin=0,ymax=0.5,
        /pgfplots/xtick pos=left,
         /pgfplots/ytick pos=left,
        /pgfplots/xtick align=center,
        /pgfplots/ytick align=center,
        xtick={1,3,5,7,9},
        ytick={0,0.1,0.2,0.3,0.4},
        width=7cm,  
        height=5cm,  
    ]

    \addplot + [only marks, mark size = 1.5pt] table [x=x, y=y, col sep=comma] {data.csv};

    \end{axis}
\end{tikzpicture}
\caption{Ordered optimal probabilities $i \mapsto p^\ast_{(i)}$\\
{\small Notes: This figure plots the argmin $p^\ast$ of \eqref{eq:VPU3}, where the $p^{\ast}_{(i)}$'s are coordinates of the vector $p^{\ast}$ in descending order. 
In Theorem~\ref{th:VP-prudent}, it is proven that the VP DM is ambiguity prudent when the map $i\mapsto p^\ast_{(i)}$ is convex. 
This plot is an illustration of a convex map that makes the DM ambiguity prudent.}}
\label{fig:convexprob}
\end{figure}

The subdifferential of a function $c$ is defined as follows:
\[\partial c(p)=\{x\in \R^k: \forall p_0\in \Delta, c(p_0)-c(p)\geq (p-p_0)x\}.\]
We state the following two theorems, describing a sufficient condition for ambiguity prudence in terms of the ambiguity index.
\begin{theorem} \label{thm: subdif convexity}
Grant Assumption~\ref{ass:sym}.        
A VP DM is ambiguity prudent if
\[\{p\in \Delta: \partial c(p)\neq \emptyset\} \subseteq  \bigg\{(.,p_1,p_2,p_3,.)|p_{(2)}\leq \frac{p_{(1)}+ p_{(3)}}{2}\bigg\}.\]
\end{theorem}

\vskip 0.2cm 
Suppose that $c(p):\Delta\to\mathbb{R}^+_0$ can be extended to a convex, differentiable function $c:[0,1]^n\to\mathbb{R}$. Denote by 
$$\bigg(\frac{\partial c}{\partial p}\bigg)^{-1}(x):=\{p\mbox{ is a probability measure }| x=\frac{\partial c}{\partial p}(p)\}.$$
Further define 
$$f(x):=\bigg(\frac{\partial c}{\partial p}\bigg)_2^{-1}(x+\lambda(1,\ldots,1)),$$
with the subscript $_2$ denoting the second largest component among the components $p_1,p_2$ and $p_3$ of the vector $p$, and with $\lambda\in\mathbb{R}$ being the unique number such that
$$(\frac{\partial c}{\partial p})^{-1}(x+\lambda(1,\ldots,1))\neq \emptyset.$$
\begin{theorem}
\label{prudentpenalty}
Grant Assumption~\ref{ass:sym}.        
A VP DM is ambiguity prudent if 
\begin{equation*}
\label{convexity}    
f(.,y_1,y_2,y_3,.)\leq  \frac{1}{2}f\left(.,\frac{y_1+y_3}{2},y_1,y_3,.\right)+\frac{1}{2}f\left(.,y_1,y_3,\frac{y_1+y_3}{2},.\right),
\end{equation*}
for $y_1\leq y_2\leq y_3,$ where this inequality should be understood for any element of the corresponding set.
\end{theorem}

\subsection{Divergence Preferences}

A special case of variational preferences is given by the so-called variational divergence preferences involving a \emph{$g$-divergence} ambiguity index; see \cite{MMR06}. 
In this subsection, we investigate this subclass. 
Divergence preferences imply probabilistic sophistication in the sense of \cite{MS92}. 
Specifically, let $g:[0,\infty)\to \R \cup \{\infty\}$ be a lower-semicontinuous, convex function satisfying $g(1)<\infty$ and $\lim_{x\to \infty}g(x)/x = \infty$. 
The $g$-divergence of two probability measures $P$ and $Q$ is given by
\begin{equation*}
I_g(Q|P):=\E\sqb{g\of{\frac{\diff Q}{\diff P}}}.
\end{equation*}
A VP DM with $g$-divergence then uses the following evaluation for an act $X$ designated in monetary units:
\[U_g(X)= \min_{Q \in \Delta} \{\E_Q[u(X)]+I_g(Q|P)\}=\min_{q\in \Delta} \{\sum_{i=1}^{k}q_{i}u(x_{i})+I_g(q|p)\},\]
where $P$ is the reference probability measure. 


Theorem~4.122 in \cite{FS16} yields that, for an act $X$,
\begin{equation}\label{eq:dualconjugate}
    U_g(X)=\sup_{z\in \R} \of{z-\E_P\sqb{g^\ast(z-u(X))}}=\sup_{z\in \R} \of{z+\E_P\sqb{f(u(X)-z)}},
\end{equation} 
with $g^*(z)=\sup_{x\geq 0}\{zx-g(x)\}$ 
the Fenchel dual conjugate and $f(z):=-g^*(-z)$ increasing and concave. 
Representations of the form \eqref{eq:dualconjugate} were introduced in \cite{BT86,BT87} as ``optimized certainty equivalents'' with the interpretation that a DM maximizes his/her utility from consumption over two time periods.
Eqn.~\eqref{eq:dualconjugate} can also be interpreted as a two period consumption problem, in which $z$ represents the utils induced by what is consumed today and is therefore not affected by ambiguity, whereas $u(X)-z$ represents the utils induced by what is consumed tomorrow and is subject to ambiguity. 
The increasing function $f$ then represents the ambiguity attitude of the DM, with $f$ being concave meaning that the DM values a gain in expected utility due to model misspecification less than a loss, and consequently prefers less variation induced by ambiguity to more.


\subsubsection{Ambiguity Prudence}
Any VP DM is ambiguity averse.
In this subsection, we characterize ambiguity prudence for variational divergence preferences.
Recall Definition~\ref{def:ambprud}.
We state the following theorem:

\begin{theorem}\label{thm:prudence with g}
Grant Assumption~\ref{ass:sym}. 
A VP DM with $g$-divergence ambiguity index is ambiguity prudent if and only if $(g^{\ast})'''\geq 0$.
\end{theorem}

We remark that $(g^*)'''\geq 0$ is equivalent to $f'''(z)=(g^*)'''(-z)\geq 0$. 
In light of \eqref{eq:dualconjugate}, ambiguity prudence therefore corresponds to a function $f$ that becomes more ambiguity averse the lower the expected utility. 
In particular, if $z$ represents the expected utility in the first period, then for an ambiguity prudent DM, an increase in the variation of $u(X)$ will, through the equation $1=\E_P[f'(X)]-z^*$ with $z^*$ being the argmax in \eqref{eq:dualconjugate} and Jensen's inequality, lead to an increase of $z^*$. 
That is, the total evaluation will correspond to a situation in which more utils are induced by early consumption and fewer by later consumption.

\renewcommand\arraystretch{2}
\begin{table}[t!]
{\footnotesize 
\begin{center}
\begin{tabular}{|c|c|c|c|}
\hline  
\textbf{Divergence}  & $g(p)$, $p\geq 0$ & $g^\ast(z)$& $(g^\ast)'''(z)$ \\
\hline \hline
Relative Entropy & $p\log(p)$ & $\exp(z-1)$& $\exp(z-1)$ \\
\hline
Burg Entropy & $p-\log(p)-1$ & $-\log(1-s)$, $s<1$& $2(1-s)^{-3}$ \\
\hline
$\chi^2$-distance & $\frac{1}{p}(p-1)^2$ & $2-2\sqrt{1-s}$, $s<1$&$\frac{3}{4}(1-s)^{-5/2}$ \\
\hline
Hellinger distance & $(\sqrt{p}-1)^2$ & $\frac{s}{1-s}$, $s<1$ & $6(1-s)^{-4}$ \\
\hline
Cressie and Read & $\frac{1-\theta+\theta t-t^\theta}{\theta(1-\theta)}$& $\frac{1}{\theta}(1-s(1-\theta))^{\frac{\theta}{\theta-1}}-\frac{1}{\theta}$, $s<\frac{1}{1-\theta}$& $(2-\theta)(1-s(1-\theta))^{\frac{3-2\theta}{\theta-1}}$\\
\hline \hline
\end{tabular}
\caption{Examples of $g$-divergences, dual conjugates and third order derivatives.\newline
{\small \textit{Notes}: This table displays some well-known $g$-divergences from the literature. 
Theorem~\ref{thm:prudence with g} provides a characterization of ambiguity prudence in terms of the non-negativity of the third derivative of the dual conjugate function $g^{\ast}$. 
For different $g$-divergence functions, the corresponding dual conjugate functions and their third derivatives are provided in this table.}}
\label{tab:1}
\end{center}}
\end{table}

\vskip 0.2cm 
Tables~2 and~4 in \cite{BHWMR13} provide some well-known $g$-divergence functions and their dual conjugates. 
The third derivatives of these dual conjugates have been calculated in Table~\ref{tab:1}. 
Hence, Theorem~\ref{thm:prudence with g} implies that a DM who uses the relative entropy, Burg entropy, $\chi^2$-distance or Hellinger distance is ambiguity prudent since $(g^\ast)'''\geq 0$ for each one of them. 

The Cressie and Read divergence constitutes a broad class of divergence functions parameterized by $\theta$. 
For $\theta=1$, it corresponds to the relative entropy; for $\theta =1/2$, it corresponds to the Hellinger distance; for $\theta=-1$, it corresponds to the $\chi^2$-distance. 
When $\theta\leq 2$, a DM who uses the Cressie and Read divergence function is ambiguity prudent.

\section{Maxmin Expected Utility}\label{sec: MaxMin}

A related canonical theory for decision under risk and ambiguity is given by the \emph{maxmin expected utility} (MEU) model, also referred to as multiple priors. 
MEU, introduced by \cite{GS89}, admits, for an act $X$ designated in monetary units, a numerical representation given by 
\begin{equation}\label{eq: maxminpref}
	U(X)=\min_{Q\in M}\mathbb{E}_{Q}[u(X)]=\min_{Q\in M}\sum_{i=1}^{k}q_{i}u(x_{i}),
\end{equation} where $M \subset\Delta:=\{p\in \mathbb{R}^{k}_{0+}\vert \sum_{i=1}^k p_i=1\}$. 
By replacing $M$ with its convex closure, we may under MEU actually always assume that the set $M$ is closed and convex. 
This also justifies writing ``min'' instead of ``inf'' in \eqref{eq: maxminpref}.
Upon setting $c(p)$ in the VP model equal to an indicator function that is zero if $p\in M$ and infinity else (i.e., $c=I_M$ for a closed convex set $M$), we can see that MEU occurs as a special case of VP, so that the results of the previous section apply, \textit{mutatis mutandis}. 

\subsection{Ambiguity Aversion}
In this subsection, we succinctly apply the results of the previous section to characterize ambiguity aversion within the MEU model. 
Recall Definition~\ref{def:ambavr}.
For 
\[u(X)=[.,E;y_1,\omega_1; y_2,\omega_2;..,E^{c}\setminus\{\omega_{1},\omega_{2}\}]\] with $y_1\leq y_2$, 
\begin{equation}
\label{eq:maxminU}
U(X)=\min_{(.,p_1,p_2,.)\in M}\{.+p_1 y_1+p_2 y_2+.\},  
\end{equation}
for an arbitrary closed and convex set $M$. 
Denote the argmin of \eqref{eq:maxminU} by $$p^*(y_1,y_2)=(.,p^*_1(y_1,y_2),p^*_2(y_1,y_2),.).$$
In the case the argmin is not unique, denote by $p^*$ an arbitrary $p$ that attains the minimum.

The following theorem is a consequence of Theorem~\ref{th:VPaa} and Remark~\ref{rem:VPaa}.
\begin{theorem}
Grant Assumption~\ref{ass:sym}.        
An MEU DM is ambiguity averse if and only if 
\begin{equation}
    \label{eq:statement3}
p^*_2(y_1,y_2)\leq  p^*_1(y_1,y_2),
\end{equation}
for any argmin $p^*$.
Furthermore, for an MEU DM, \eqref{eq:statement3} holds for any argmin. 
In particular, an MEU DM is always ambiguity averse.
\end{theorem}

\subsection{Ambiguity Prudence}
Next, we analyze the implications of ambiguity prudence within the MEU model.
Recall Definition~\ref{def:ambprud}.
For \[u(X)=[.,E;y_1,\omega_1; y_2,\omega_2;y_3,\omega_3;..,E^{c}\setminus\{\omega_{1},\omega_{2},\omega_{3}\}]\] with $y_1\leq y_2\leq y_3$, 
\begin{equation}
    \label{maxminU2}
    U(X)=\min_{(.,p_1,p_2,p_3,.)\in M}\{.+p_1 y_1+p_2 y_2+p_3 y_3+.\},
\end{equation}  
for an arbitrary closed and convex set $M$.
Denote the argmin of \eqref{maxminU2} by $$p^*(y_1,y_2,y_3)=(.,p^*_1(y_1,y_2,y_3),p^*_2(y_1,y_2,y_3),p^*_3(y_1,y_2,y_3),.).$$ 

We state the following theorems, which follow from Theorems~\ref{th:VP-prudent-necessary} and~\ref{th:VP-prudent}:
\begin{theorem}
Grant Assumption~\ref{ass:sym}.        
An MEU DM is ambiguity prudent only if 
\begin{equation*}
p^*_2(y_1,y_2,y_3)\leq  \frac{p^*_1(y_1,y_2,y_3)+p^*_3(y_1,y_2,y_3)}{2},
\end{equation*}
for any argmin $p^*$ whenever $y_2-y_1 >y_3-y_2$ with $y_1\leq y_2\leq y_3$.
\end{theorem}

\begin{theorem}\label{th:maxminprudent}
Grant Assumption~\ref{ass:sym}.        
An MEU DM is ambiguity prudent if 
\begin{equation}
    \label{statement}
p^*_2(y_1,y_2,y_3)\leq  \frac{p^*_1(y_1,y_2,y_3)+p^*_3(y_1,y_2,y_3)}{2},
\end{equation}
for any argmin $p^*.$
\end{theorem}

Note that we can recover the case of CEU within the MEU model.
As is well-known (\citealp{S86,S89}), the CEU model occurs as a special case of the MEU model provided the capacity is supermodular.
However, not all MEU models can be expressed as a CEU model with a supermodular capacity. 
Recall that all MEU models are ambiguity averse. 
We now translate the ambiguity prudence characterization of the CEU model to its counterpart within the MEU model.
Specifically, identifying in \eqref{eq:upau} 
$\nu(E^c)$ with $d+p_1+p_2+p_3$, 
$\nu(E^c)\setminus \{{\omega_1}\}$ with $d+p_2+p_3$, and $\nu(E^c)\setminus \{\omega_1,\omega_2\}$ with $d+p_3$ for a constant $d$, 
we obtain from \eqref{eq:upau} that $p^*_1-2p^*_2+p^*_3\geq 0$. 
In particular, $p^*_2\leq\frac{p^*_1+p^*_3}{2}$.

The next theorem provides a characterization of the structure of the set $M$ when the MEU DM is ambiguity prudent. 
Denote by $\mbox{relint}(M)$ the relative interior of the set $M$ and, by $\partial_{rel} M$, $M$ without its relative interior.
\begin{theorem}\label{th:rel}
If we choose $M$ to be closed and convex, then the MEU DM is ambiguity prudent if 
\begin{align}
\label{partialM}
&\partial_{rel} M\cap \{p \,|U(y)= p y \text{ for a }  y\text{ with }y_i\neq y_j\text{ for }i\neq j \}\nonumber\\
&\hspace{1cm}\subset  \bigg\{(.,p_1,p_2,p_3,.)|p_{(2)}\leq \frac{p_{(1)}+ p_{(3)}}{2}\bigg\} .
    \end{align} 
\end{theorem}

\begin{remark}
    If there are no non-trivial restrictions on the probabilities associated to the three relevant states, condition~\eqref{partialM} is always satisfied; all probability mass of the three states is then shifted to the worst state. 
    In this case, the set $M$ is given by the set of all probabilities $p$ and 
    $$M=conv(\{e^1,e^2,\ldots \}),$$
    with $e_i$ defined as the unit vector being one in the $i$-th component and zero everywhere else, and $conv$ denotes the convex hull. 
    In this example, we can also see directly that \eqref{partialM} holds, as
    $$\partial_{rel}  M\cap \{p \,|U(y)= p y \text{ for a }  y\text{ with }y_i\neq y_j\text{ for }i\neq j \}= \left\{e_1,e_2,\ldots\right\},$$
    and
    $$e^{i}_{(2)}=0<\frac{1}{2}=\frac{e^i_{(1)}+e^i_{(3)}}{2}, \mbox{ for each }i.$$
\end{remark}

\begin{remark}
\label{counter}
    It is not true that ambiguity prudence implies that 
    \begin{equation*}
    \partial_{rel}  M \subset  \bigg\{(.,p_1,p_2,p_3,.)|p_{(2)}\leq \frac{p_{(1)}+ p_{(3)}}{2}\bigg\},
    \end{equation*} 
    with $\partial_{rel}M$ defined as above (i.e., as the boundary of the relative closure of the relative interior).
    For instance, by the previous remark, in the case that there are no non-trivial restrictions on the probabilities associated to the three relevant states, we have that the MEU DM is ambiguity prudent, and the closure of the relative interior of $M$ is given by $$\{z| z_i=0\mbox{ for at least one }i\}.$$ 
    Thus, $p=(1/2,1/2,0,\ldots,0)$ is in the boundary of the relative interior, $\partial_{rel}M$, but $p_{(2)}\leq \frac{p_{(1)}+ p_{(3)}}{2}$ actually does not hold for this $p$.
\end{remark} 

The next corollary follows immediately from Theorem~\ref{th:maxminprudent}.
\begin{corollary}
Suppose that $M$ is the polytope generated by (corners) $p_1,\ldots,p_k$, i.e., $$M=conv(p^1,\ldots,p^k),$$ 
and $p_1,\ldots,p_k$ are linearly independent. 
Then the MEU DM is ambiguity prudent if each $p^j$ satisfies \eqref{statement} for $j=1,\ldots,k$.
\end{corollary}
For examples of polytopes of alternative probabilistic models, see \cite{BRS21}, \cite{KLLSS18}, \cite{LSSS24} or \cite{CVW23}.

\subsection{Multiple Priors with Level-\texorpdfstring{$K$}{\textit{K}} Priors}

In this subsection, we consider a special but quite general set of priors $M_{f,K}:=\{P: \sum_{i=1}^k f(p_i)\leq K\}$ for a lower-semicontinuous, proper, convex function $f:[0,\infty) \to \R \cup \{\infty\}$ satisfying $f(1)<\infty$, $f(0)=0$, and $K\geq 0$ such that $\{P: \sum_{i=1}^k f(p_i)< K\}\neq \emptyset$. 
Note that $M_{f,K}$ is convex, closed and bounded, hence it is compact. 
 
\subsubsection{Ambiguity Prudence}
Any MEU DM is ambiguity averse.
We now analyze ambiguity prudence for multiple priors with level-$K$ priors.
Recall Definition~\ref{def:ambprud}.
We state the following theorem:

\begin{theorem} \label{thm: prudence for bounded MP}
Grant Assumption~\ref{ass:sym}. 
An MEU DM with the set of priors $M_{f,K}$ is ambiguity prudent if $f(1)\leq K$. 
If $f(1)>K$, then the MEU DM is ambiguity prudent if $(f^\ast)'''\geq 0$, where $f^{\ast}(x)=\sup_{q\geq 0}(qx-f(q))$.
\end{theorem}

\begin{remark}
For a reference probability measure $P$, one may desire to determine the set of priors by constructing a `bounded ball' around this reference measure. 
Assumption~\ref{ass:sym} gives that the reference probability measure is uniform. 
For a function $f$, one can describe this ball as $\cb{Q \in \Delta: \sum_{i=1}^{k} f\of{\frac{q_i}{1/k}}\leq K}$. 
As is clear from the definition, this setting is within the range of Theorem~\ref{thm: prudence for bounded MP}.  
For instance, for the $p$-norm, we have $f(q)=\frac{1}{p}q^p$. 
Therefore, a DM who uses the $p$-norm for $1\leq p\leq 2$ is ambiguity prudent, since $f^\ast(x)=\frac{p-1}{p}x^{\frac{p}{p-1}}$ for $x\geq 0$ with $(f^\ast)'''(x)=\frac{2-p}{(p-1)^2}x^{\frac{3-2p}{p-1}}\geq 0$.
\end{remark}
\begin{example}
Suppose that the set of priors consists of probability measures that have bounded relative entropy, i.e., $f(q)=q\log(q)$. 
Then, the DM is ambiguity prudent, since $f^\ast(x)=\exp(x-1)$. 
For a discussion and applications of this set, see \cite{FK11} and \cite{PS20}.
\end{example}
\begin{example}
A DM who uses the Hellinger distance $f(q)=(\sqrt{q}-1)^2-1$ 
is ambiguity prudent. 
\end{example}

\subsection{\texorpdfstring{$\alpha$}{alpha}-Maxmin}

A related theory is given by the \textit{$\alpha$-maxmin} model.
The numerical representation of the $\alpha$-maxmin model, for an act $X$ designated in monetary units, is given by (\citealp{M02}, \citealp{GMM04}, \citealp{CEG07}, \citealp{GP14})
\begin{align*}
	U(X)=\alpha\min_{Q\in M}\mathbb{E}_Q[u(X)]+(1-\alpha)\max_{Q\in M}\mathbb{E}_Q[u(X)], \quad \alpha\in [0,1],
\end{align*}
for a set of priors $M$.
A popular special case of interest includes the Hurwicz expected utility model (\citealp{GP14,GP15}).

In this subsection, we consider a special case of the $\alpha$-maxmin model that is referred to as $\epsilon$-contamination, introduced in \cite{EK99}. 
In this case, the set of priors $M$ is given by 
\begin{equation*} 
M=\{Q: Q=(1-\epsilon)P+\epsilon R, R\in \Delta\},\quad \epsilon\in [0,1],
\end{equation*}
where $P$ is a reference probability measure and $\Delta$ is the set of all probability distributions. 
Here, $\epsilon$ corresponds to the level of skepticism about the probabilistic benchmark $P$.

\subsubsection{Ambiguity Aversion}
We state the following result:
\begin{theorem}\label{thm: aversion for epsilon}
Grant Assumption~\ref{ass:sym}. 
An $\epsilon$-contamination DM is ambiguity averse if and only if $\alpha=1$.
\end{theorem}

\subsubsection{Ambiguity Prudence}
We state the following theorem:
\begin{theorem}\label{thm: prudence for epsilon}
Grant Assumption~\ref{ass:sym}. 
An $\epsilon$-contamination DM is ambiguity prudent.
\end{theorem}

\vskip 0.2 cm

Note that when $\alpha = 1$, the $\alpha$-maxmin model reduces to the multiple priors model. 
Therefore, Theorems~\ref{thm: aversion for epsilon} and~\ref{thm: prudence for epsilon} imply that the multiple priors model is ambiguity averse and ambiguity prudent when the set of priors is given by $M=\{Q:Q=(1-\epsilon)P+\epsilon R, R\in \Delta\}$.

\section{Smooth Ambiguity Model}\label{sec:SA}

In this section, we consider the \emph{Smooth Ambiguity} (SA) model developed by \cite{KMM05}; see also \cite{DP22}.
Denote by $M$ a set of countably additive probability measures on $(S, \Sigma)$. 
Suppose $M$ is endowed with the weak$^\ast$ topology and the corresponding Borel $\sigma$-algebra.
Then the SA model admits, for an act $X$ designated in monetary units, a numerical representation of the form
\begin{equation*}
U(X)=\mathbb{E}_\mu[\phi(\mathbb{E}_P[u(X)])]=\int_{M}\phi\left(\int_{S}u(X)\diff P\right)\diff \mu(P),
\end{equation*}
where $\mu$ is a non-atomic prior on $M$,  $\phi:\mathbb{R}\rightarrow\mathbb{R}$ is an increasing function, capturing ambiguity attitudes, and $u:\mathbb{R}\rightarrow\mathbb{R}$ is a utility function. 
Henceforth, we impose the following assumption.
\begin{assumption}\label{Assum: SA}
    If $(.,p_1,p_2,.)\in M$, then $(.,p_2,p_1,.)\in M$.
\end{assumption}

We state the following lemmas.
\begin{lemma}\label{lem:IndifferenceforSA}
Assumption~\ref{ass:sym} holds if $\mu$ is symmetric, i.e., $\mu(.,p_1,p_2,.)=\mu(.,p_2,p_1,.)$ for all $p_1,p_2$.
\end{lemma}


\begin{lemma}\label{lem: phi symm}
Suppose that $\varphi \in C^\infty(\mathbb{R})$ is concave, not linear and strictly increasing. If Assumption~\ref{ass:sym} holds, then $\mu$ is symmetric, i.e., $\mu(.,p_1,p_2,.)=\mu(.,p_2,p_1,.)$ for all $p_1,p_2$.
\end{lemma}

\begin{lemma}\label{lemma1-add}
Let $\tilde{\mu}_1$ and $\tilde{\mu}_2$ be two distributions on $\mathbb{R}^n$ (or $\mathbb{R}^\infty$) with support restricted to the unit simplex (i.e., with support on non-negative numbers adding up to one).
Suppose that $\varphi \in C^\infty(\mathbb{R})$ is concave, not linear and strictly increasing.  
If
\begin{equation}\label{eq: eq1}
\mathbb{E}[\varphi(xP_1)]=\mathbb{E}[\varphi(xP_2)] \text{ for all } x,
\end{equation} with $P_i\stackrel{d}{=}\tilde{\mu}_i \text{ for } i=1,2$, then $\tilde{\mu}_1=\tilde{\mu}_2$.
\end{lemma}

\begin{remark}
If $\varphi$ is polynomial of order $k<n-1$, clearly Eqn.~\eqref{eq: eq1} implies at most that the first $k$ moments of $\tilde{\mu}_1$ and $\tilde{\mu}_2$ agree, which is not sufficient to conclude that $\tilde{\mu}_1=\tilde{\mu}_2$.
\end{remark}

\subsection{Second Order Expected Utility}

We proceed in this subsection by considering the special case of the second order expected utility model, for which the treatment simplifies.

The second order expected utility model (SOEU) has been introduced by \cite{N93,N10}. 
The SOEU model admits the following representation:
\begin{equation*}\label{eq: SOEU}
    U(X)=\int_{S} \phi(u(X(\omega)))\diff P(\omega).
\end{equation*}
\begin{lemma}\label{lem: SOEU sym}
    Assumption~\ref{ass:sym} implies that the probability distribution is uniform.
\end{lemma}

The next theorem shows that within SOEU our definition of ambiguity aversion coincides with the one given by \cite{KMM05}.
\begin{theorem}\label{thm: SOEU aversion}
Grant Assumption~\ref{ass:sym}. 
An SOEU DM is ambiguity averse if and only if $\phi$ is concave.
\end{theorem}

The following theorem characterizes ambiguity prudence within SOEU.

\begin{theorem}\label{thm: SOEU prudence}
Grant Assumption~\ref{ass:sym}. 
An SOEU DM is ambiguity prudent if and only if $\phi'''\geq 0$.
\end{theorem}

By ambiguity aversion, the DM is more concerned about a possible loss of one util than enticed by the possible gain of one util. 
Theorem~\ref{thm: SOEU prudence} shows that ambiguity prudence means that this difference gets larger the smaller the utils.

\subsection{Ambiguity Aversion}
In this subsection, we characterize ambiguity aversion within the SA model.
We state the following theorem.

\begin{theorem}\label{thm: SA aversion}
Suppose that $\mu$ is symmetric. 
Then, an SA DM is ambiguity averse if and only if $\phi$ is concave.
\end{theorem}

\subsection{Ambiguity Prudence}
Next, we characterize ambiguity prudence within the SA model.

\begin{theorem} \label{thm: SA prudence}
Suppose that $M \subseteq \cb{(.,p_1,p_2,p_3,.)|p_{(2)}\leq \frac{p_{(1)}+ p_{(3)}}{2}}$ and $\mu$ is symmetric.
Then, an SA DM is ambiguity prudent if and only if $\phi'''\geq 0$.
\end{theorem}

Note that within SA $\phi$ evaluates the expected utilities of different models, where each model is weighted by $\mu$ according to its perceived plausibility. 
Hence, the concavity of $\phi$ implies that an additional unit of expected utility due to model misspecification is less valuable the wealthier the DM already is. 
This is equivalent to a potential loss of expected utility due to model missspecification hurting the DM more than a potential gain of expected utility. 
It is instructive to compare this to the effect of a concave utility function $u$. 
Here, the loss of one Euro due to risk (i.e., due to randomness stemming from the \emph{known} underlying probability distribution of a lottery) hurts the DM more than the additional gain of one Euro gives additional satisfaction. 
We remark that for studying the shape and the derivative of $\phi$, the effect of decreasing marginal utility due to the DM's attitude toward wealth is already taken care of as $\phi$ evaluates expected utilities rather than payoffs.
Hence, the variation in the expected utilities (i.e., losses and gains of expected utilities) entering the evaluation through $\phi$ is entirely due to potential model missspecification and the DM's aversion to it.\footnote{In particular, a DM who is ambiguity neutral (i.e., uses a linear $\phi$) is of course indifferent between the potential gain of expected utility or its potential loss due to using an incorrect probability distribution.}

\begin{theorem} \label{thm: SA prudence 2}
Suppose that $\phi'''\geq 0$ and $\mu$ is symmetric. 
Then, an SA DM is ambiguity prudent if and only if 
$M \subseteq \cb{(.,p_1,p_2,p_3,.)|p_{(2)}\leq \frac{p_{(1)}+p_{(3)}}{2}}$.
\end{theorem}

\begin{remark}
These two theorems give the following: 
Suppose that a SA DM is ambiguity prudent. 
Then, $\phi'''\geq 0$ if and only if $M \subseteq \cb{(.,p_1,p_2,p_3,.)|p_{(2)}\leq \frac{p_{(1)}+ p_{(3)}}{2}}$.
\end{remark}

\section{
Insuring an Ambiguous Loss}\label{sec: Insurance g-div}

In this section, we study an optimal insurance problem.
We show that for a DM with variational preferences and ambiguity index represented by a $g$-divergence, the optimal policy is explicitly linked to ambiguity prudence.
The insurance problem we analyze is somewhat similar to that considered in \cite{EJT12}.
These authors consider three `policies': no insurance, full insurance, and partial insurance with a fixed indemnity $s$ and premium $\pi$, to cover a loss $\ell$. 
We investigate a more general model in which the insurance premium varies with the indemnity, which takes values in a continuum. 
Later, we will also allow for uncertainty with respect to the loss amount. 

Let us formalize the problem.
Consider a VP DM with an affine utility function and ambiguity index represented by a $g$-divergence. 
The generalization to a non-affine utility function is discussed in Section~\ref{sec:nonaffine} and the extension to the SOEU model is considered in Section~\ref{sec:appSOEU}, yielding similar results.
She has initial wealth $w>0$ and faces a loss $\ell>0$ with unknown loss probability. 
She considers insuring herself against the loss. 
For an indemnity equal to $s$, $0\leq s\leq \ell$, the insurance premium equals $\pi(s)$, with $\pi(\cdot)$ increasing, convex and continuously differentiable. 
The DM aims to find the optimal insurance policy and solves:
\begin{equation}
\label{eq: linear g-div example}
    \max_{0\leq s \leq\ell}\min_{Q\in \Delta}\left\{ \E_Q[X(s)]+I_g(Q|P)\right\},
\end{equation}
where
\begin{equation*}
X(s):= [w-\ell+s-\pi(s),\omega_1;w-\pi(s),\omega_2;\ldots;w-\pi(s),\omega_k].
\end{equation*}
Under Assumption~\ref{ass:sym}, using Theorem~4.122 in \cite{FS16}, we can re-write \eqref{eq: linear g-div example} as
\begin{equation*}
\max_{0\leq s\leq\ell, z\in\R}\left\{z-\frac{1}{k}g^\ast(z-w+\ell-s+\pi(s))-\frac{k-1}{k}g^\ast(z-w+\pi(s))\right\}.
\end{equation*}
Applying the change of variable $\tilde{z}:=z-w+\pi(s)$ transforms the problem into:
\begin{equation*}
w+\max_{0\leq s\leq\ell, \tilde{z}\in\R}\left\{\tilde{z}-\pi(s)-\frac{1}{k}g^\ast(\tilde{z}+\ell-s)-\frac{k-1}{k}g^\ast(\tilde{z})\right\}.
\end{equation*}
Denote the optimal solutions by $\tilde{z}^\ast$ and $s^\ast$. 
They need to satisfy the respective first-order conditions. 
The first-order condition with respect to $\tilde{z}$ gives
\begin{equation}\label{eq: FOC for z without risk}
    1=\frac{1}{k}(g^\ast)'(\tilde{z}^\ast+\ell-s^\ast)+\frac{k-1}{k}(g^\ast)'(\tilde{z}^\ast),
\end{equation}
whereas the first-order condition with respect to $s$ yields that
\begin{equation}\label{eq: FOC for s without risk}
\pi'(s^\ast)=\frac{1}{k}(g^\ast)'(\tilde{z}^\ast+\ell-s^\ast)=1-\frac{k-1}{k}(g^\ast)'(\tilde{z}^\ast). 
\end{equation}

Now, additionally suppose that there is `noise' with respect to the loss amount, i.e., the loss is either $\ell-\epsilon$ or $\ell+\epsilon$, $\epsilon>0$, with unknown probabilities. 
Then, the DM faces the following problem:
\begin{equation}\label{eq: linear g-div example with noise}
    \max_{0\leq s\leq\ell}\min_{Q\in \Delta} \left\{\E_Q[X_\epsilon(s)]+I_g(Q|P)\right\},
\end{equation}
where
\[X_\epsilon(s):=[w-\ell-\epsilon+s-\pi(s),\omega_{(1,\epsilon)};w-\ell+\epsilon+s-\pi(s),\omega_{(1,-\epsilon)}; w-\pi(s),\omega_2;\ldots;w-\pi(s),\omega_k].\]
Under Assumption~\ref{ass:sym}, using Theorem 4.122 in \cite{FS16}, we can re-write \eqref{eq: linear g-div example with noise} as
\begin{align*}
\max_{0\leq s\leq\ell,z\in \R}&\left\{z-\frac{1}{2k}g^\ast(z-w+\ell+\epsilon-s+\pi(s))\right.\\
&\left.-\frac{1}{2k}g^\ast(z-w+\ell-\epsilon-s+\pi(s))-\frac{k-1}{k}g^\ast(z-w+\pi(s))\right\}.
\end{align*}
Applying again the change of variable $\tilde{z}=z-w+\pi(s)$ transforms the problem into:
\begin{equation*}
w+\max_{s\geq 0,\tilde{z}\in \R}\left\{\tilde{z}-\pi(s)-\frac{1}{2k}g^\ast(\tilde{z}+\ell+\epsilon-s)-\frac{1}{2k}g^\ast(\tilde{z}+\ell-\epsilon-s)-\frac{k-1}{k}g^\ast(\tilde{z})\right\}.
\end{equation*}
Denote the optimal solutions of this problem by $\tilde{z}^\ast_\epsilon$ and $s^\ast_\epsilon$. 
The first-order condition with respect to $\tilde{z}$ gives
\begin{equation}\label{eq: FOC for z with risk}
 1=\frac{1}{2k}(g^\ast)'(\tilde{z}^\ast_\epsilon+\ell+\epsilon-s^\ast_\epsilon)+\frac{1}{2k}(g^\ast)'(\tilde{z}^\ast_\epsilon+\ell-\epsilon-s^\ast_\epsilon)+\frac{k-1}{k}(g^\ast)'(\tilde{z}^\ast_\epsilon),
\end{equation}
whereas the first-order condition with respect to $s$ yields that
\begin{equation}\label{eq: FOC for s with risk}
\pi'(s^\ast_\epsilon)=\frac{1}{2k}(g^\ast)'(\tilde{z}^\ast_\epsilon+\ell+\epsilon-s^\ast_\epsilon)+\frac{1}{2k}(g^\ast)'(\tilde{z}^\ast_\epsilon+\ell-\epsilon-s^\ast_\epsilon)=1-\frac{k-1}{k}(g^\ast)'(\tilde{z}^\ast_\epsilon). 
\end{equation}

To understand how the presence of noise with respect to the loss amount influences the optimal insurance decision of the DM, let us compare \eqref{eq: FOC for z without risk} and \eqref{eq: FOC for z with risk}. 
Note that, by our main results, ambiguity prudence, i.e., $(g^\ast)'''\geq 0$, implies the convexity of $(g^\ast)'$, which gives that
\begin{align*}
 1&=\frac{1}{k}(g^\ast)'(\tilde{z}^\ast+\ell-s^\ast)+\frac{k-1}{k}(g^\ast)'(\tilde{z}^\ast)
 \\&=\frac{1}{2k}(g^\ast)'(\tilde{z}^\ast_\epsilon+\ell+\epsilon-s^\ast_\epsilon)+\frac{1}{2k}(g^\ast)'(\tilde{z}^\ast_\epsilon+\ell-\epsilon-s^\ast_\epsilon)+\frac{k-1}{k}(g^\ast)'(\tilde{z}^\ast_\epsilon)
 \\& \geq \frac{1}{k}(g^\ast)'(\tilde{z}^\ast_\epsilon+\ell-s^\ast_\epsilon)+\frac{k-1}{k}(g^\ast)'(\tilde{z}^\ast_\epsilon),
\end{align*}
i.e.,
\begin{equation}\label{eq:app-pivotal}
    \frac{1}{k}(g^\ast)'(\tilde{z}^\ast+\ell-s^\ast)+\frac{k-1}{k}(g^\ast)'(\tilde{z}^\ast) \geq \frac{1}{k}(g^\ast)'(\tilde{z}^\ast_\epsilon+\ell-s^\ast_\epsilon)+\frac{k-1}{k}(g^\ast)'(\tilde{z}^\ast_\epsilon).
\end{equation}
By definition, $g^\ast$ is convex, which yields the increasingness of $(g^\ast)'$. 
Therefore, \eqref{eq:app-pivotal} implies one of the following:
\begin{itemize}
    \item[(i)] $\tilde{z}^\ast \geq \tilde{z}^\ast_\epsilon$,
    \item [(ii)] if $\tilde{z}^\ast \leq \tilde{z}^\ast_\epsilon$, then $\tilde{z}^\ast-s^\ast\geq \tilde{z}^\ast_\epsilon -s^\ast_\epsilon$.
\end{itemize}
If (ii) is the case, it directly follows that $s^\ast_\epsilon \geq s^\ast$. 
Let us investigate (i), using \eqref{eq: FOC for s without risk} and \eqref{eq: FOC for s with risk}. 
Using $\tilde{z}^\ast \geq \tilde{z}^\ast_\epsilon$ and the monotonicity of $(g^\ast)'$, we have
\begin{align*}
    \pi'(s^\ast)=1-\frac{k-1}{k}(g^\ast)'(\tilde{z}^\ast)\leq 1-\frac{k-1}{k}(g^\ast)'(\tilde{z}^\ast_\epsilon)=\pi'(s^\ast_\epsilon).
\end{align*}
Therefore, the convexity of $\pi$ gives that $s_\epsilon^\ast \geq s^\ast$. 
It can be concluded that in the presence of uncertainty with respect to the loss amount, \textit{ceteris paribus} the prudent DM wishes to purchase more insurance.

\section{Conclusion}\label{sec:con}

In this paper, we have introduced an ambiguity preference, as a primitive trait of behavior.
Its one-shot and double-shot application yield simple definitions of ambiguity aversion and ambiguity prudence 
that admit easy interpretations.
We have formally established the implications of these model-free definitions of ambiguity aversion and ambiguity prudence within canonical models for decision under risk and ambiguity, including Choquet Expected Utility, Variational Preferences and Smooth Ambiguity.
In particular, we have provided intuitive economic interpretations of the higher-order derivatives of capacities. 
We have also shown that ambiguity prudence can be naturally connected to an optimal insurance problem under ambiguity.
Because our definitions are simple, they are amenable to experimental testing and verification. 
We intend to pursue this in future work.

\appendix 
\renewcommand{\theequation}{A.\arabic{equation}}
\setcounter{equation}{0}

\section{Appendix}

\subsection{Proofs of Section~\ref{sec:CEU}}
\begin{proof}[Proof of Theorem~\ref{th:CEU2nd-i}]
A non-negative second derivative of the capacity means
\begin{align}
&\nu(A\cup\{\omega_{1},\omega_{2}\})-\nu(A\cup\{\omega_{1}\}\setminus\{\omega_{2}\})\nonumber\\
&-\nu(A\cup\{\omega_{2}\}\setminus\{\omega_{1}\})+\nu(A\setminus\{\omega_{1},\omega_{2}\})
\geq 0,\label{eq:pos2ndder}
\end{align}
for any $A\subseteq S$.

The CEU evaluation $V(X_{A}^{(2)}(\omega_{1},\omega_{2}))$ is given by
\begin{align*}
&.\nu(S)+(u_{1}-\bar{u}-.)\nu(E^{c})
+(u_{2}+\bar{u}-(u_{1}-\bar{u}))\nu(E^{c}\setminus\{\omega_{1}\})\\
&+(..-(u_{2}+\bar{u}))\nu(E^{c}\setminus\{\omega_{1},\omega_{2}\}).
\end{align*}
Similarly, $V(X_{B}^{(2)}(\omega_{1},\omega_{2}))$ is given by
\begin{align*}
&.\nu(S)+(u_{1}+\bar{u}-.)\nu(E^{c})
+(u_{2}-\bar{u}-(u_{1}+\bar{u}))\nu(E^{c}\setminus\{\omega_{1}\})\\
&+(..-(u_{2}-\bar{u}))\nu(E^{c}\setminus\{\omega_{1},\omega_{2}\}).
\end{align*}

Hence, $V(X_{B}^{(2)}(\omega_{1},\omega_{2}))-V(X_{A}^{(2)}(\omega_{1},\omega_{2}))$, i.e., the utility premium under the CEU model, takes the form
\begin{align*}
2\left(\bar{u}\nu(E^{c})
-2\bar{u}\nu(E^{c}\setminus\{\omega_{1}\})
+\bar{u}\nu(E^{c}\setminus\{\omega_{1},\omega_{2}\})\right).
\end{align*}
Similarly, $V(X_{B}^{(2)}(\omega_{1},\omega_{2}))-V(X_{A}^{(2)}(\omega_{1},\omega_{2}))$ takes the form
\begin{align*}
2\left(\bar{u}\nu(E^{c})
-2\bar{u}\nu(E^{c}\setminus\{\omega_{2}\})
+\bar{u}\nu(E^{c}\setminus\{\omega_{1},\omega_{2}\})\right).
\end{align*}
If we sum these two expressions, we have
\begin{align*}
    4\bar{u}(\nu(E^{c})-
\nu(E^{c}\setminus\{\omega_{1}\})-\nu(E^{c}\setminus\{\omega_{2}\})
+\nu(E^{c}\setminus\{\omega_{1},\omega_{2}\})).
\end{align*}
By the arbitrariness of $E$, the preference of $X_{B}^{(2)}$ to $X_{A}^{(2)}$ for any permutation of $\omega_{1},\omega_{2}$, 
then implies that the second derivative of the capacity $\nu$ is non-negative.
\end{proof}
\vskip 0.2cm
\begin{proof}[Proof of Theorem~\ref{th:CEU2nd-ii}]
Recall~\eqref{eq:pos2ndder}.
Hence,
\begin{align*}
&V(X_{B}^{(2)}(\omega_{1},\omega_{2}))
-V(X_{A}^{(2)}(\omega_{1},\omega_{2}))\\
&+V(X_{B}^{(2)}(\omega_{2},\omega_{1}))
-V(X_{A}^{(2)}(\omega_{2},\omega_{1}))\geq 0.
\end{align*}
As, by Assumption~\ref{ass:sym}, 
\begin{equation*}
V(X_{B}^{(2)}(\omega_{1},\omega_{2}))
=V(X_{B}^{(2)}(\omega_{2},\omega_{1})),
\end{equation*}
and 
\begin{equation*}
V(X_{A}^{(2)}(\omega_{1},\omega_{2}))
=V(X_{A}^{(2)}(\omega_{2},\omega_{1})),
\end{equation*}
the stated result follows.
\end{proof}

\vskip 0.2cm

\begin{proof}[Proof of Theorem~\ref{th:CEU3rd-i}]
A non-negative third derivative of the capacity means
\begin{align}
&\nu(A\cup\{\omega_{1},\omega_{2},\omega_{3}\})\nonumber\\
&-\nu(A\cup\{\omega_{1},\omega_{2}\}\setminus\{\omega_{3}\})-\nu(A\cup\{\omega_{1},\omega_{3}\}\setminus\{\omega_{2}\})-\nu(A\cup\{\omega_{2},\omega_{3}\}\setminus\{\omega_{1}\})\nonumber\\
&+\nu(A\cup\{\omega_{1}\}\setminus\{\omega_{2},\omega_{3}\})+\nu(A\cup\{\omega_{2}\}\setminus\{\omega_{1},\omega_{3}\})+\nu(A\cup\{\omega_{3}\}\setminus\{\omega_{1},\omega_{2}\})\nonumber\\
&-\nu(A\setminus\{\omega_{1},\omega_{2},\omega_{3}\})
\geq 0,\label{eq:pos3rdder}
\end{align}
for any $A\subseteq S$.

The CEU evaluation $V(X_{A}^{(3)}(\omega_{1},\omega_{2},\omega_{3}))$ is given by
\begin{align*}
&.\nu(S)+(u_{1}-\bar{u}-.)\nu(E^{c})
+(u_{2}+2\bar{u}-(u_{1}-\bar{u}))\nu(E^{c}\setminus\{\omega_{1}\})\\
&+(u_{3}-\bar{u}-(u_{2}+2\bar{u}))\nu(E^{c}\setminus\{\omega_{1},\omega_{2}\})
+(..-(u_{3}-\bar{u}))\nu(E^{c}\setminus\{\omega_{1},\omega_{2},\omega_{3}\}).
\end{align*}
Similarly, $V(X_{B}^{(3)}(\omega_{1},\omega_{2},\omega_{3}))$ is given by
\begin{align*}
&.\nu(S)+(u_{1}+\bar{u}-.)\nu(E^{c})
+(u_{2}-2\bar{u}-(u_{1}+\bar{u}))\nu(E^{c}\setminus\{\omega_{1}\})\\
&+(u_{3}+\bar{u}-(u_{2}-2\bar{u}))\nu(E^{c}\setminus\{\omega_{1},\omega_{2}\})
+(..-(u_{3}+\bar{u}))\nu(E^{c}\setminus\{\omega_{1},\omega_{2},\omega_{3}\}).
\end{align*}

Hence, $V(X_{B}^{(3)}(\omega_{1},\omega_{2},\omega_{3}))-V(X_{A}^{(3)}(\omega_{1},\omega_{2},\omega_{3}))$, i.e., the utility premium under the CEU model, takes the form
\begin{align}\label{eq:upau}
2\left(\bar{u}\nu(E^{c})
-3\bar{u}\nu(E^{c}\setminus\{\omega_{1}\})
+3\bar{u}\nu(E^{c}\setminus\{\omega_{1},\omega_{2}\})
-\bar{u}\nu(E^{c}\setminus\{\omega_{1},\omega_{2},\omega_{3}\})\right)\geq 0.
\end{align}
Similarly, $V(X_{B}^{(3)}(\omega_{3},\omega_{1},\omega_{2}))-V(X_{A}^{(3)}(\omega_{3},\omega_{1},\omega_{2}))$ takes the form 
\begin{align*}
2\left(\bar{u}\nu(E^{c})
-3\bar{u}\nu(E^{c}\setminus\{\omega_{3}\})
+3\bar{u}\nu(E^{c}\setminus\{\omega_{1},\omega_{3}\})
-\bar{u}\nu(E^{c}\setminus\{\omega_{1},\omega_{2},\omega_{3}\})\right)\geq 0.
\end{align*}
and $V(X_{B}^{(3)}(\omega_{2},\omega_{3},\omega_{1}))-V(X_{A}^{(3)}(\omega_{2},\omega_{3},\omega_{1}))$ taks the form
\begin{align*}
2\left(\bar{u}\nu(E^{c})
-3\bar{u}\nu(E^{c}\setminus\{\omega_{2}\})
+3\bar{u}\nu(E^{c}\setminus\{\omega_{2},\omega_{3}\})
-\bar{u}\nu(E^{c}\setminus\{\omega_{1},\omega_{2},\omega_{3}\})\right)\geq 0.
\end{align*}
If we sum these three expressions, we obtain
\begin{align*}
    6\bar{u}\big(&\nu(E^c)-\nu(E^{c}\setminus\{\omega_{1}\})-\nu(E^{c}\setminus\{\omega_{2}\})-\nu(E^{c}\setminus\{\omega_{3}\})\\&+\nu(E^{c}\setminus\{\omega_{1},\omega_{2}\})+\nu(E^{c}\setminus\{\omega_{2},\omega_{3}\})+\nu(E^{c}\setminus\{\omega_{1},\omega_{3}\})-\nu(E^{c}\setminus\{\omega_{1},\omega_{2},\omega_{3}\}) \big) \geq 0
\end{align*}
By the arbitrariness of $E$, the preference of $X_{B}^{(3)}$ to $X_{A}^{(3)}$ for any permutation of $\omega_{1},\omega_{2},\omega_{3}$,
then implies that
the third derivative of the capacity $\nu$ is non-negative.
\end{proof}

\vskip 0.2cm

\begin{proof}[Proof of Theorem~\ref{th:CEU3rd-ii}]
Recall~\eqref{eq:pos3rdder}.
Hence,
\begin{align*}
&V(X_{B}^{(3)}(\omega_{1},\omega_{2},\omega_{3}))
-V(X_{A}^{(3)}(\omega_{1},\omega_{2},\omega_{3}))\\
&+V(X_{B}^{(3)}(\omega_{2},\omega_{3},\omega_{1}))
-V(X_{A}^{(3)}(\omega_{2},\omega_{3},\omega_{1}))\\
&+V(X_{B}^{(3)}(\omega_{3},\omega_{1},\omega_{2}))
-V(X_{A}^{(3)}(\omega_{3},\omega_{1},\omega_{2}))\geq 0.
\end{align*}
As, by Assumption~\ref{ass:sym}, 
\begin{equation*}
V(X_{B}^{(3)}(\omega_{1},\omega_{2},\omega_{3}))
=V(X_{B}^{(3)}(\omega_{2},\omega_{3},\omega_{1}))
=V(X_{B}^{(3)}(\omega_{3},\omega_{1},\omega_{2})),
\end{equation*}
and 
\begin{equation*}
V(X_{A}^{(3)}(\omega_{1},\omega_{2},\omega_{3}))
=V(X_{A}^{(3)}(\omega_{2},\omega_{3},\omega_{1}))
=V(X_{A}^{(3)}(\omega_{3},\omega_{1},\omega_{2})),
\end{equation*}
the stated result follows.
\end{proof}

\vskip 0.2cm

\begin{proof}[Proof of Theorem~\ref{thm:ambavr-NCEU}]
The second derivative of the capacity $\nu$ for a set $E$ with respect to $\omega_1$ and $\omega_2$ is given by:
\begin{align*}
\nu_{\omega_1,\omega_2}(E)=\nu(E\cup\{\omega_{1},\omega_{2}\})-\nu(E\cup\{\omega_{1}\}\setminus\{\omega_{2}\})-\nu(E\cup\{\omega_{2}\}\setminus\{\omega_{1}\})+\nu(E\setminus\{\omega_{1},\omega_{2}\}).
\end{align*}
Let us calculate the second derivative of the neo-additive capacity for any $\omega_1,\omega_2$ and a set $E$. 
We investigate four cases.
\begin{itemize}
    \item [(i)] Suppose that $S$ has only two elements, i.e., $S=\{\omega_1,\omega_2\}$. 
    Then, for any $E\subseteq S$, we have
    \begin{align*}
      \nu_{\omega_1,\omega_2}(E)&=1-((a-b)/2+(1-a)P(\omega_1))-((a-b)/2+(1-a)P(\omega_2))+0 
      \\&=1-2(a-b)/2-(1-a)(P(\omega_1)+P(\omega_2))=1-(a-b)-(1-a)=b.
    \end{align*}
    \item [(ii)] Suppose that $S$ has more than two elements and $E\setminus \{\omega_1,\omega_2\}=\emptyset$. 
    Then, we have
    \begin{align*}
        \nu_{\omega_1,\omega_2}(E)&=(a-b)/2+(1-a)(P(\omega_1)+P(\omega_2))
        \\&-((a-b)/2+(1-a)P(\omega_1))-((a-b)/2+(1-a)P(\omega_2))+0
        \\&=(a-b)/2-2(a-b)/2+(1-a)(P(\omega_1)+P(\omega_2)-P(\omega_1)-P(\omega_2))
        \\&=(b-a)/2.
    \end{align*}
    \item [(iii)] Suppose that $S$ has more than two elements and $E\cup \{\omega_1,\omega_2\}=S$. 
    Then, we have
    \begin{align*}
      \nu_{\omega_1,\omega_2}(E)&=1 -((a-b)/2+(1-a)(1-P(\omega_2)))-((a-b)/2+(1-a)(1-P(\omega_1)))      
      \\&+(a-b)/2+(1-a)(1-P(\omega_1)-P(\omega_2))
      \\&=1-2(a-b)/2+(a-b)/2-(1-a)
      \\&+(1-a)(P(\omega_1)+P(\omega_2)-P(\omega_1)-P(\omega_2))
      \\&=a-(a-b)/2=(a+b)/2.
    \end{align*}
    \item [(iv)] For the other cases, since 
    $\nu_{\omega_1,\omega_2}(E)=\nu_{\omega_1,\omega_2}(E\setminus \{\omega_1,\omega_2\}),$ without loss of generality suppose that $E\cap\{\omega_1,\omega_2\}=\emptyset$. 
    Then, we have
    \begin{align*}
       \nu_{\omega_1,\omega_2}(E)&=(a-b)/2+(1-a)(P(E)+P(\omega_1)+P(\omega_2))
       \\&-((a-b)/2+(1-a)(P(E)+P(\omega_2))) 
       \\&-((a-b)/2+(1-a)(P(E)+P(\omega_1)))+((a-b)/2+(1-a)P(E))
       \\&=2(a-b)/2-2(a-b)/2
       \\&+(1-a)(2P(E)-2P(E)+P(\omega_1)+P(\omega_2)-P(\omega_1)-P(\omega_2))
       \\&=0.
    \end{align*}
\end{itemize} 
Theorems~\ref{th:CEU2nd-i} and~\ref{th:CEU2nd-ii} imply that a CEU DM is ambiguity averse if and only if the second derivative of the capacity is non-negative. 
Combining this result with (i), (ii), (iii), and (iv), we obtain that a NCEU DM is ambiguity averse if and only if $b\geq 0$, $(b-a)/2 \geq 0$, $(a+b)/2 \geq 0$. 
Note that we have $a\in[0,1]$ and $b\in [-a,a]$ by the definition of the NCEU model. 
Hence, it follows that the DM is ambiguity averse if and only if $a=b$ and $b\geq 0$. 
Note that $b\geq 0$ is redundant since $a=b$ and $a\geq 0$ by definition.
\end{proof}

\vskip 0.2cm
\begin{proof}[Proof of Theorem~\ref{thm:ambprud-NCEU}]
The third derivative of the capacity $\nu$ at $E$ with respect to $\omega_1,\omega_2, \omega_3$ is given by:
\begin{align*}
&\Delta_{\omega_{1},\omega_{2},\omega_{3}}\nu(E)\\
&=\nu(E\cup\{\omega_{1},\omega_{2},\omega_{3}\})\\
&\quad-\nu(E\cup\{\omega_{1},\omega_{2}\}\setminus\{\omega_{3}\})-\nu(E\cup\{\omega_{1},\omega_{3}\}\setminus\{\omega_{2}\})-\nu(E\cup\{\omega_{2},\omega_{3}\}\setminus\{\omega_{1}\})\\
&\quad+\nu(E\cup\{\omega_{1}\}\setminus\{\omega_{2},\omega_{3}\})+\nu(E\cup\{\omega_{2}\}\setminus\{\omega_{1},\omega_{3}\})+\nu(E\cup\{\omega_{3}\}\setminus\{\omega_{1},\omega_{2}\})\\
&\quad-\nu(E\setminus\{\omega_{1},\omega_{2},\omega_{3}\}).
\end{align*}
Let us calculate the third derivative of the neo-additive capacity for any $\omega_1,\omega_2,\omega_3$ and a set $E$. 
We investigate four cases.
\begin{itemize}
\item [(i)] Suppose that $S$ has only three elements, i.e., $S=\{\omega_1,\omega_2,\omega_3\}.$ 
Then, for any $E\subseteq S$, we have
\begin{align*}
    \Delta_{\omega_{1},\omega_{2},\omega_{3}}\nu(E)&=1-((a-b)/2+(1-a)(P(\omega_1)+P(\omega_2)))
    \\&-((a-b)/2+(1-a)(P(\omega_1)+P(\omega_3)))
    \\&-((a-b)/2+(1-a)(P(\omega_2)+P(\omega_3)))
    +(a-b)/2+(1-a)P(\omega_1)
\\&+(a-b)/2+(1-a)P(\omega_2)+(a-b)/2+(1-a)P(\omega_3)-0
\\&=1-3(a-b)/2+3(a-b)/2
\\&-(1-a)((P(\omega_1)+P(\omega_2))+(P(\omega_2)+P(\omega_3))+(P(\omega_1)+P(\omega_3)))
\\&+(1-a)(P(\omega_1)+P(\omega_2)+P(\omega_3))-0
\\&=1-(1-a)(P(\omega_1)+P(\omega_2)+P(\omega_3))=a\geq 0,
\end{align*}
since $a\in[0,1]$.
\item [(ii)] Suppose that $S$ has more than three elements and take $E\setminus \{\omega_1,\omega_2,\omega_3\}=\emptyset$. 
Then, we have
\begin{align*}
\Delta_{\omega_{1},\omega_{2},\omega_{3}}\nu(E)&=(a-b)/2+(1-a)(P(\omega_1)+P(\omega_2)+P(\omega_3))
\\&-((a-b)/2+(1-a)(P(\omega_1)+P(\omega_2)))
\\&-((a-b)/2+(1-a)(P(\omega_1)+P(\omega_3)))
\\&-((a-b)/2+(1-a)(P(\omega_2)+P(\omega_3)))
+(a-b)/2+(1-a)P(\omega_1)
\\&+(a-b)/2+(1-a)P(\omega_2)+(a-b)/2+(1-a)P(\omega_3)-0
\\&= 4(a-b)/2-3(a-b)/2
+(1-a)(P(\omega_1)+P(\omega_2)+P(\omega_3))
\\&-(1-a)(P(\omega_1)+P(\omega_2)+P(\omega_1)+P(\omega_3)+P(\omega_2)+P(\omega_3))
\\&+(1-a)(P(\omega_1)+P(\omega_2)+P(\omega_3))
\\&=(a-b)/2\geq 0,
\end{align*}
since $b\in[-a,a]$. 
\item [(iii)] Suppose that $S$ has more than three elements and take  $E\cup \{\omega_1,\omega_2,\omega_3\} =S$. 
Then, we have
\begin{align*}
\Delta_{\omega_{1},\omega_{2},\omega_{3}}\nu(E) &= 1 -((a-b)/2+(1-a)(1-P(\omega_3)))
\\&-((a-b)/2+(1-a)(1-P(\omega_2)))
-((a-b)/2+(1-a)(1-P(\omega_1)))
\\&+((a-b)/2+(1-a)(1-P(\omega_2)-P(\omega_3)))
\\&+((a-b)/2+(1-a)(1-P(\omega_1)-P(\omega_3)))
\\&+((a-b)/2+(1-a)(1-P(\omega_1)-P(\omega_2)))
\\&-((a-b)/2+(1-a)(1-P(\omega_1)-P(\omega_2)-P(\omega_3)))
\\&=1-(a-b)/2-(1-a)(1-P(\omega_3)+1-P(\omega_2)+1-P(\omega_1))
\\&+(1-a)(1-P(\omega_2)-P(\omega_3))+(1-a)(1-P(\omega_1)-P(\omega_3))
\\&+(1-a)(1-P(\omega_1)-P(\omega_2))-(1-a)(1-P(\omega_1)-P(\omega_2)-P(\omega_3))
\\&=1-(a-b)/2 -(1-a)=(a+b)/2 \geq 0,
\end{align*}
since $b\in[-a,a]$.
\item [(iv)] For the other cases, since 
    $\nu_{\omega_1,\omega_2,\omega_3}(E)=\nu_{\omega_1,\omega_2,\omega_3}(E\setminus \{\omega_1,\omega_2,\omega_3\})$, without loss of generality suppose that $E\cap \{\omega_1,\omega_2,\omega_3\}=\emptyset.$ 
Then, we have
\begin{align*}
\Delta_{\omega_{1},\omega_{2},\omega_{3}}\nu(E)&=   (a-b)/2+(1-a)(P(E)+P(\omega_1)+P(\omega_2)+P(\omega_3))
\\&-((a-b)/2+(1-a)(P(E)+P(\omega_1)+P(\omega_2)))
\\&-((a-b)/2+(1-a)(P(E)+P(\omega_1)+P(\omega_3)))\\&-((a-b)/2+(1-a)(P(E)+P(\omega_2)+P(\omega_3)))
\\&+(a-b)/2+(1-a)(P(E)+P(\omega_1))
\\&+(a-b)/2+(1-a)(P(E)+P(\omega_2))
\\&+(a-b)/2+(1-a)(P(E)+P(\omega_3))
-((a-b)/2+(1-a)P(E))
\\&=4(a-b)/2-4(a-b)/2+(1-a)(4P(E)-4P(E))
\\&+2(1-a)(P(\omega_1)+P(\omega_2)+p(\omega_3))
-2(1-a)(P(\omega_1)+P(\omega_2)+p(\omega_3))
\\&=0.\end{align*}
\end{itemize}
Therefore, we conclude that the third derivative is always non-negative for the neo-additive CEU model. 
Hence, a NCEU DM is always ambiguity prudent as a consequence of Theorems~\ref{th:CEU3rd-i} and~\ref{th:CEU3rd-ii}.
\end{proof}
\subsection{Proofs of Section~\ref{sec:Var}}
\begin{proof}[Proof of Theorem~\ref{thm: indifference-sym}]
Recall \eqref{eq:X_I^2}.
It is sufficient to show that a VP DM is indifferent between any $X_{I}^{(2)}(\omega_{1},\omega_{2})$ and $X_{I}^{(2)}(\omega_{2},\omega_{1})$ if and only if the ambiguity index satisfies the conditions of the theorem.
It is easy to see that if $c$ is symmetric, then the VP DM is indifferent between any $X_{I}^{(2)}(\omega_{1},\omega_{2})$ and $X_{I}^{(2)}(\omega_{2},\omega_{1})$. 
To prove the converse, suppose that $U(X_{I}^{(2)}(\omega_{1},\omega_{2}))=U(X_{I}^{(2)}(\omega_{2},\omega_{1}))$ for any $\omega_1,\omega_2$ and $E$. 
First, note that $c$ can be written as follows:
      \[c(p)=\sup_{X\in \mathcal{F}}\{U(X)-\E_p[u(X)]\},\]
by invoking Theorem~3 and Proposition~6 in \cite{MMR06}. 
Next, choose probability measures $p^1, p^2$ such that $p^1(\omega_1)=p_1$ and $p^1(\omega_2)=p_2$, and $p^2(\omega_1)=p_2$ and $p^2(\omega_2)=p_1$. 
Then, we have
      \begin{equation*}\label{eq: sym of x1 and x2}
      U(X_{I}^{(2)}(\omega_{1},\omega_{2}))-\E_{p^1}[u(X_{I}^{(2)}(\omega_{1},\omega_{2}))]=U(X_{I}^{(2)}(\omega_{2},\omega_{1}))-\E_{p^2}[u(X_{I}^{(2)}(\omega_{2},\omega_{1}))].
      \end{equation*} 
This is true for any $X_{I}^{(2)}(\omega_{1},\omega_{2})$ and $X_{I}^{(2)}(\omega_{2},\omega_{1})$, hence
      \begin{align*}
      c(p^1)&=\sup_{X_{I}^{(2)}(\omega_{1},\omega_{2})\in \mathcal{F}} \{U(X_{I}^{(2)}(\omega_{1},\omega_{2}))-\E_{p^1}[u(X_{I}^{(2)}(\omega_{1},\omega_{2}))]\}
      \\&=\sup_{X_{I}^{(2)}(\omega_{2},\omega_{1})\in \mathcal{F}} \{U(X_{I}^{(2)}(\omega_{2},\omega_{1}))-\E_{p^2}[u(X_{I}^{(2)}(\omega_{2},\omega_{1}))]\}=c(p^2),
      \end{align*}
where the second equality is valid because, for any $u(X_{I}^{(2)}(\omega_{1},\omega_{2})) \in \mathcal{F}$, there is a corresponding $u(X_{I}^{(2)}(\omega_{2},\omega_{1}))\in \mathcal{F}$, and \textit{vice versa}. 
Because of the arbitrariness of the choice of $p^1$ and $p^2$, the result follows.
\end{proof}
\vskip 0.2cm 

\begin{proof}[Proof of Theorem~\ref{th:VPaa}]
First, assume that \eqref{eq:statementAAforVP} does not hold. 
Then,
\begin{align*}
U(X_A^{(2)})=&\ .+p_1^{\ast}(u_1-\bar{u},u_2+\bar{u})u_1+p_2^{\ast}(u_1-\bar{u},u_2+\bar{u})u_2+.\\
&+c(.,p_1^*(u_1-\bar{u},u_2+\bar{u}),p_2^*(u_1-\bar{u},u_2+\bar{u}),.)\\
&-\bar{u}(p_1^{\ast}(u_1-\bar{u},u_2+\bar{u})-p_2^{\ast}(u_1-\bar{u},u_2+\bar{u}))\\
>&\ .+p_1^{\ast}(u_1-\bar{u},u_2+\bar{u})u_1+p_2^{\ast}(u_1-\bar{u},u_2+\bar{u})u_2+.\\
&+c(.,p_1^*(u_1-\bar{u},u_2+\bar{u}),p_2^*(u_1-\bar{u},u_2+\bar{u}),.)\\
&+\bar{u}(p_1^{\ast}(u_1-\bar{u},u_2+\bar{u})-p_2^{\ast}(u_1-\bar{u},u_2+\bar{u}))\\
\geq&\ U(X_B^{(2)}),
\end{align*}
where the first inequality follows from the positivity of $\bar{u}$ and the last one is a consequence of the definition of variational preferences.

Next, to prove the reverse direction, suppose that \eqref{eq:statementAAforVP} holds for all $y_1\leq y_2$.
Then, we have
\begin{align*}
U(X_B^{(2)})&=\min_{(.,p_1,p_2,.) \in \Delta, p_2\leq p_1}\{.+p_1(u_1+\bar{u})+p_2(u_2-\bar{u})+.+c(.,p_1,p_2,.)\}
\\&=\min_{(.,p_1,p_2,.) \in \Delta, p_2\leq p_1}\{.+p_1u_1+p_2u_2+\bar{u}(p_1-p_2)+.+c(.,p_1,p_2,.)\}
\\&\geq \min_{(.,p_1,p_2,.) \in \Delta, p_2\leq p_1}\{.+p_1u_1+p_2u_2-\bar{u} (p_1-p_2)+.+c(.,p_1,p_2,.)\}
=U(X_A^{(2)}),
\end{align*}
which proves the stated result.
\end{proof}

\vskip 0.2cm

\begin{proof}[Proof of Theorem~\ref{th:VP-prudent-necessary}]
Fix a $p^*$. 
First of all, note that \eqref{eq:con} is equivalent to 
$$p^*_2(y_1,y_2,y_3)\leq \frac{p^*_1(y_1,y_2,y_3)+p^*_2(y_1,y_2,y_3)+p^*_3(y_1,y_2,y_3)}{3}=:\frac{q^*}{3},$$
where $y_2-y_1 > y_3-y_2$.

Let us assume by contradiction that \eqref{eq:con} does not hold for some $y_1,y_2,y_3$ satisfying $y_2-y_1 > y_3-y_2$. 
Denote $(.,p_1,p_2,p_3,.)$ by $\bar{p}$ for brevity.
Then,
\begin{align*}
    &U(X_B^{(3)})
    \\&\leq\min_{\bar{p}\in \Delta,\, p_2> \frac{p_1+p_2+p_3}{3}}\{.+p_1 (u_1+\bar{u})+p_2 (u_2-2\bar{u})+p_3 (u_3+\bar{u})+\ldots +c(.,p_1,p_2,p_3,.)\}\\
    &=\min_{\bar{p}\in \Delta,\,p_2> \frac{q}{3},\, q=p_1+p_2+p_3 }\{.+p_1 (u_1+\bar{u})+p_2 (u_2-2\bar{u})\\
    & \hspace{4cm}+(q-p_1-p_2)(u_3+\bar{u})+\ldots +c(.,p_1,p_2,p_3,.)\}\\
    &=\min_{q\in[0,1]}\bigg\{2q\bar{u}+\min_{\bar{p}\in \Delta,\, \,p_2> \frac{q}{3},\,q=p_1+p_2+p_3}\{.+p_1 (u_1-\bar{u})+p_2 (u_2-4\bar{u})\\
    &\hspace{5.2cm}+(q-p_1-p_2) (u_3-\bar{u})+\ldots +c(.,p_1,p_2,p_3,.)\}
   \bigg\} 
    \\&<\min_{q\in[0,1]}\bigg\{2q\bar{u}-2q\bar{u} +\min_{\bar{p}\in \Delta,\, p_2> \frac{q}{3},\, q=p_1+p_2+p_3}\{.+p_1 (u_1-\bar{u})+p_2 (u_2+2\bar{u})\\
    &\hspace{5.5cm}+p_3 (u_3-\bar{u})+\ldots +c(.,p_1,p_2,p_3,.)\}\bigg\}\\
    & =\min_{\bar{p}\in \Delta,\, p_2> 
    \frac{p_1+p_2+p_3}{3}}\{.+p_1 (u_1-\bar{u})+p_2 (u_2+2\bar{u})+p_3 (u_3-\bar{u})+\ldots +c(.,p_1,p_2,p_3,.)\}\\
    &=U(X_A^{(3)}),
\end{align*}
where the first inequality holds by definition of $U$ and the last equality holds by violation of \eqref{eq:con}, with $u_1,\bar{u},\Delta u$ such that $u_1-\bar{u}=y_1$, $u_2+2\bar{u}=y_2$, $u_3-\bar{u}=y_3$. 
This inequality is a contradiction to the DM being prudent. 
Hence, if the DM is prudent, indeed \eqref{eq:con} holds.
\end{proof}

\vskip 0.2 cm

\begin{proof}[Proof of Theorem~\ref{th:VP-prudent}]
Suppose that \eqref{eq:statement for VP} holds. 
Denote $(.,p_1,p_2,p_3,.)$ by $\bar{p}$ for brevity.
Then, we have
    \begin{align*}
           U(X_B^{(3)})&=\min_{\bar{p}\in \Delta,p_2\leq \frac{p_1+p_3}{2}} (.+p_1(u_1+\bar{u})+p_2(u_2-2\bar{u})+p_3(u_3+\bar{u})+.+c(.,p_1,p_2,p_3,.))
           \\&=\min_{\bar{p}\in \Delta,p_2\leq \frac{p_1+p_3}{2}} (.+p_1u_1+p_2u_2+p_3u_3+.+c(.,p_1,p_2,p_3,.)+\bar{u}(p_1-2p_2+p_3))
           \\&\geq \min_{\bar{p}\in \Delta,p_2\leq \frac{p_1+p_3}{2}} (.+p_1u_1+p_2u_2+p_3u_3+.+c(.,p_1,p_2,p_3,.)-\bar{u}(p_1-2p_2+p_3))
           \\&=U(X_A^{(3)}).
    \end{align*}
\end{proof}

\vskip 0.2 cm
\begin{proof}[Proof of Theorem~\ref{thm: subdif convexity}]
Suppose that 
  \[\{p\in \Delta: \partial c(p)\neq \emptyset\} \subseteq  \bigg\{(.,p_1,p_2,p_3,.)|p_{(2)}\leq \frac{p_{(1)}+ p_{(3)}}{2}\bigg\}.\]
From the definition of the subdifferential, we have that $p^{\ast}$ is an argmin of \eqref{eq:VPU3} if and only if $x\in \partial c(p^\ast)$. 
This gives $\partial c(p^{\ast})\neq \emptyset$ and $p^{\ast}_{(2)}\leq \frac{p^{\ast}_{(1)}+p^{\ast}_{(3)}}{2}$. 
Therefore, the VP DM is ambiguity prudent because of Theorem~\ref{th:VP-prudent}.
    %
\end{proof}

\vskip 0.2 cm

\begin{proof}[Proof of Theorem~\ref{prudentpenalty}]
Let us analyze the minimization problem \eqref{eq:VPU3}. 
Since by lower-semicontinuity of $c$ a minimum exists, Lagrange optimization yields that \eqref{eq:VPU3} is equivalent to 
$$\argmin_p\bigg\{pu-c(p)-\lambda(1-\sum_i p_i)-\sum_i \mu_i p_i\bigg\}=\argmin_p\bigg\{pu-c(p)-\lambda(1-\sum_i p_i)\bigg\},$$ since by assumption $p^*>0$ so that the constraints to which the $\mu_i$'s relate are non-binding. 

Note that, by the definition of the subgradient, $p^*$ being the $\argmin$ is equivalent to $$y+\lambda(1,\ldots,1)\in \bigg(\frac{\partial c}{\partial p}\bigg) (p^*) .$$ 
This again is equivalent to $$p^*\in\bigg(\frac{\partial c}{\partial p}\bigg)^{-1}(y+\lambda(1,\ldots,1)).$$
Since by Theorem~\ref{thm: indifference-sym} $c$ is symmetric, we have that $$\frac{\partial c (\ldots,p_i,\ldots,p_j,\ldots)}{\partial p_i}=\frac{\partial c(\ldots,p_j,\ldots,p_i,\ldots)}{\partial p_j},$$ 
implying that, for $(\ldots,y_i+\lambda,\ldots,y_j+\lambda,\ldots)$, 
we have 
$$\bigg(\frac{\partial c}{\partial p}\bigg)_i^{-1}(\ldots,y_i+\lambda,\ldots,y_j+\lambda,\ldots)=\bigg(\frac{\partial c}{\partial p}\bigg)_j^{-1}(\ldots,y_j+\lambda,\ldots,y_i+\lambda,\ldots).$$
Considering $y_1<y_3$ and $y_2=(y_1+y_3)/2$, condition~\eqref{eq:statement for VP} is then equivalent to
\begin{align*}
&\bigg(\frac{\partial c}{\partial p}\bigg)_2^{-1}\bigg(.,y_1+\lambda,\frac{y_1+y_3}{2}+\lambda,y_3+\lambda,.\bigg)\\
\hspace{0.1cm}
&\leq \frac{1}{2} \bigg(\frac{\partial c}{\partial p}\bigg)^{-1}_1\bigg(.,y_1+\lambda,\frac{y_1+\lambda+y_3+\lambda}{2},+y_3,.\bigg)\\
&\quad+\frac{1}{2}\bigg(\frac{\partial c}{\partial p}\bigg)^{-1}_3\bigg(.,+y_1,\frac{y_1+\lambda+y_3+\lambda}{2},y_3,.\bigg)\\
&=\frac{1}{2}\bigg(\frac{\partial c}{\partial p}\bigg)_2^{-1}\bigg(.,\frac{y_1+\lambda+y_3+\lambda}{2},y_1+\lambda,y_3+ \lambda,.\bigg)\\
&\quad+\frac{1}{2}\bigg(\frac{\partial c}{\partial p}\bigg)^{-1}_2\bigg(.,y_1+\lambda,y_3+\lambda,\frac{y_1+\lambda+y_3+\lambda}{2},.\bigg),
\end{align*}
for any $y_1<y_3$, where ``$\leq$'' should hold for any element in the corresponding set. 
The last inequality yields the theorem.
\end{proof}

\vskip 0.2 cm
The following lemma will be used in several proofs so we state it separately.
\begin{lemma}\label{lem: third-derivative}
    Let $f:\R \to \R$  be three times differentiable. 
    Then $f'''\geq 0$ if and only if 
    \begin{align*}
        f(x+\delta)+f(y-2\delta)+f(z+\delta) \geq        
        f(x-\delta)+f(y+2\delta)+f(z-\delta),
    \end{align*}
    for all $x,y,z \in \R$ satisfying $x\leq y \leq z$ and $y=(x+z)/2$, and for all $0\leq \delta \leq (z-x)/2$.
\end{lemma}
\begin{proof}
Let us define the function $h:\R\to \R$
\[h_{\delta}(t):=f(t+\delta)-f(t-\delta).\]
Since $h''(t)=f''(t+\delta)-f''(t-\delta)$, $h''\geq 0$ if and only if $f''$ is increasing, which is equivalent to $f'''\geq 0$.
We can rewrite the following inequality:
    \begin{align*}
        f(x+\delta)+f(y-2\delta)+f(z+\delta) \geq        
        f(x-\delta)+f(y+2\delta)+f(z-\delta)
    \end{align*}
    as
    \begin{align*}
        f(x+\delta)-f(x-\delta)+f(z+\delta)-f(z-\delta) \geq        
    f(y+2\delta)-f(y)+f(y)-f(y-2\delta),
    \end{align*}
    that is, 
    \begin{align*}
        h_\delta(x)+h_\delta(z)\geq h_\delta(y+\delta)+h_\delta(y-\delta),
    \end{align*}
    and
    \begin{align*}
        h_\delta(y+\epsilon)+h_\delta(y-\epsilon)\geq h_\delta(y+\delta)+h_\delta(y-\delta),
    \end{align*}
    where $\epsilon:=z-y=y-x$. 
    Since this inequality holds for any $\epsilon\geq \delta$, this is equivalent to the convexity of $h_\delta$, which is equivalent to $f'''\geq 0$.
\end{proof}

\vskip 0.2 cm

\begin{proof}[Proof of Theorem~\ref{thm:prudence with g}]
For an act $X$, 
\[U_g(X)= \min_{Q \in \Delta} (\E_Q[u(X)]+I_g(Q|P)).\]
Recall \eqref{eq:dualconjugate}.
Due to Assumption~\ref{ass:sym}, the reference probability measure $P$ is the uniform distribution. 
Hence, we can write
\begin{equation*}\label{eq: g-div dual}
   U_g(X)=\sup_{z\in \R} \of{z-\frac{1}{k}\sum_{i=1}^k g^\ast(z-y_i)}
\end{equation*}
for any act $u(X)=[y_1,\omega_1;\ldots,y_k,\omega_k]$.
Now, consider the acts
\begin{equation*}
u(X_{A}^{(3)}(\omega_{1},\omega_{2},\omega_{3}))=[.,E ; u_1-\bar{u},\omega_1 ; u_2+2\bar{u},\omega_2 ; u_3-\bar{u},\omega_3 ; ..,E^{c}\setminus\{\omega_{1},\omega_{2},\omega_{3}\}]
\end{equation*}
and
\begin{equation*}
u(X_{B}^{(3)}(\omega_{1},\omega_{2},\omega_{3}))=[.,E ; u_1+\bar{u},\omega_1 ; u_2-2\bar{u},\omega_2 ; u_3+\bar{u},\omega_3 ; ..,E^{c}\setminus\{\omega_{1},\omega_{2},\omega_{3}\}].
\end{equation*}
Lemma~\ref{lem: third-derivative} yields that
\[g^\ast(z-u_1-\bar{u})+g^\ast(z-u_2+2\bar{u})+g^\ast(z-u_3-\bar{u})\leq g^\ast(z-u_1+\bar{u})+g^\ast(z-u_2-2\bar{u})+g^\ast(z-u_3+\bar{u}), \]
for any $z\in \R$. 
Therefore, we have
\begin{align*}
    &z-\frac{1}{k}\of{\sum_{i\neq\{1,2,3\}}g^\ast(z-u_i) +g^\ast(z-u_1-\bar{u})+g^\ast(z-u_2+2\bar{u})+g^\ast(z-u_3-\bar{u})}
    \\ &\geq     z-\frac{1}{k}\of{\sum_{i\neq\{1,2,3\}}g^\ast(z-u_i) +g^\ast(z-u_1+\bar{u})+g^\ast(z-u_2-2\bar{u})+g^\ast(z-u_3+\bar{u})},
\end{align*}
for all $z\in \R$. 
Taking the supremum on both sides yields $U_g\of{X^{(3)}_B}\geq U_g\of{X^{(3)}_A}$.

To prove the converse, suppose that $(g^\ast)'''\geq 0$ does not hold, i.e., $(g^\ast)'''<0$ for some interval $[a,b]$. 
Note that by choosing $c \in (a,b)$, we can define a function $\Tilde{g}^\ast (x)= g^{\ast}(x-c).$ 
Then, $(\Tilde{g}^\ast)'''(x)< 0$ for $x \in [a-c,b-c]$ since $(\Tilde{g}^\ast)'''(x)= (g^\ast)'''(x-c)$. 
Furthermore, we have \begin{align*}
    U_g(X)&=\sup_{z\in \R}(z-\E[g^\ast(z-u(X))])=\sup_{z\in \R}(z-\E[g^\ast(z+c-u(X)-c)]
    \\&=\sup_{z\in \R}(z-\E[\Tilde{g}^\ast(z+c-u(X))])=\sup_{\Tilde{z}\in \R}(\Tilde{z}-c-\E[\Tilde{g}^\ast(\Tilde{z}-u(X)])
    \\&= U_{\Tilde{g}}(X)-c,
\end{align*}
for any act $X$.
Therefore, we can see that $U_g(X)\geq U_g(Y) \iff U_{\Tilde{g}}(X)\geq U_{\Tilde{g}}(Y)$. 
As a consequence, without loss of generality we can suppose that $0\in [a,b]$. 
Now, let us choose $u_1,u_2,u_3$ and $\bar{u}$ such that $[u_1-u_3-2\bar{u},u_3-u_1+2\bar{u}]\subset [a,b]$. 
We work with the following acts:
\begin{align*}
u(X_A^{(3)})&=[u_1-\bar{u},\omega_1;u_2+2\bar{u},\omega_2;u_3-\bar{u},\omega_3],
\\
 u(X_B^{(3)})&=[u_1+\bar{u},\omega_1;u_2-2\bar{u},\omega_2;u_3+\bar{u},\omega_3].
\end{align*}
Proposition~2.1 in \cite{BT07} gives that 
\begin{align*}\label{eq: optimal g-div B}
    U_g\of{X_B^{(3)}}&=\max_{z\in [u_1+\bar{u},u_3+\bar{u}]}\of{z-\frac{1}{3}(g^\ast(z-u_1-\bar{u})+g^\ast(z-u_2+2\bar{u})+g^\ast(z-u_3-\bar{u}))}
    \\U_g\of{X_A^{(3)}}&=\max_{z\in [u_1-\bar{u},u_3-\bar{u}]}\of{z-\frac{1}{3}(g^\ast(z-u_1+\bar{u})+g^\ast(z-u_2-2\bar{u})+g^\ast(z-u_3+\bar{u}))},
\end{align*}
where maximums are attained respectively in $z_B$ and $z_A$.
Since $g^{\ast}<0$ in $[u_1-u_3-2\bar{u},u_3-u_1+2\bar{u}]\subset[a,b]$, due to Lemma~\ref{lem: third-derivative}, we have
\begin{align*}
   &g^\ast(z-u_1-\bar{u})+g^\ast(z-u_2+2\bar{u})+g^\ast(z-u_3-\bar{u})\\
   &> g^\ast(z-u_1+\bar{u})+g^\ast(z-u_2-2\bar{u})+g^\ast(z-u_3+\bar{u})
\end{align*}
for all $z\in [u_1-\bar{u},u_3+\bar{u}]$. 
Therefore, we have 
\begin{align*}
    U_g\of{X_A^{(3)}}&=\max_{z\in [u_1-\bar{u},u_3-\bar{u}]}\of{z-\frac{1}{3}(g^\ast(z-u_1+\bar{u})+g^\ast(z-u_2-2\bar{u})+g^\ast(z-u_3+\bar{u}))}
    \\&=\max_{z\in [u_1-\bar{u},u_3+\bar{u}]}\of{z-\frac{1}{3}(g^\ast(z-u_1+\bar{u})+g^\ast(z-u_2-2\bar{u})+g^\ast(z-u_3+\bar{u}))}
    \\&\geq z_B-\frac{1}{3}(g^\ast(z_B-u_1+\bar{u})+g^\ast(z_B-u_2-2\bar{u})+g^\ast(z_B-u_3+\bar{u}))
    \\&>z_B-\frac{1}{3}(g^\ast(z_B-u_1-\bar{u})+g^\ast(z_B-u_2+2\bar{u})+g^\ast(z_B-u_3-\bar{u}))
    \\&=U_g\of{X_B^{(3)}},
    \end{align*}
which contradicts ambiguity prudence.
This proves the stated result.
\end{proof}

\subsection{Proofs of Section~\ref{sec: MaxMin}}

\begin{proof}[Proof of Theorem~\ref{th:rel}]
The theorem follows directly from Theorem~\ref{th:maxminprudent} and the fact that the minimum in \eqref{maxminU2} for states with different consequences can not be attained in the relative interior of $M$, see Lemma~\ref{lemma1}. 
Then, automatically, the relative interior of a set $M$ is a subset of $M$.
\end{proof}

\vskip 0.2 cm

\begin{lemma}
\label{lemma1}
    If $p$ is in the relative interior of $M$ and $p\neq (1/k,\ldots,1/k)$, then $y_i=y_j$ for some $i$ and $j$ for the $y$ that satisfies $\argmin_{q\in M}qy=p$.
\end{lemma}
\begin{proof}
    Note that since the boundary of the set $M$ is $M$, there exists $y$ such that $$\argmin_{q\in M}qy=p$$ due to the separating hyperplane theorem, see, e.g., \citet[page 51]{BV04}. 
    Suppose that $y_i\neq y_j$ for all $i,j$. 
    Without loss of generality, assume that $y_1<y_2<\ldots<y_{k}$. 
    Due to the minimality of $py$, we have $p_1\geq p_2\geq \ldots \geq p_{k}$. 
    There exists $i$ and $j$ such that $p_i>p_j$. 
    Take $\bar{p}\in M$ such that $\bar{p}_{l}=p_{l}$ for all $l\neq i,j$, $\bar{p}_i=p_j$, and $\bar{p}_j=p_i$. 
    Then, we have $py<\bar{p}y$. 
    Since $p$ is in the relative interior of $M$, there exist $\lambda \in (0,1)$ and $r\in M$ such that $p=\lambda \bar{p}+(1-\lambda)r$. 
    Then, by using the minimality of $p$, we have $py\leq ry.$ 
    By using these inequalities, we have
    \[py<\lambda \bar{p}y+(1-\lambda)ry=py,\]
    which is a contradiction. 
    Therefore, for some $i$ and $j$, $y_i=y_j$ holds.
\end{proof}

\vskip 0.2 cm

\begin{proof} [Proof of Theorem~\ref{thm: prudence for bounded MP}]
For the first part, note that $f(1)\leq K$ makes $(1,0,\ldots,0)$ feasible. 
Therefore, we have $U(X)=\min_{q\in M_{f,K}}\sum_{i=1}^k q_iy_i=\min(y_1,\ldots,y_k)$ for any act $X$. 
Hence, we obtain $U\of{X^{(3)}_B}\geq U\of{X^{(3)}_A}$ since $u_1+\bar{u}>u_1-\bar{u}$.

Now, suppose that $f(1)>K$.
We can rewrite the problem $\min_{q \in M_{f,K}}\sum_{i=1}^{k} q_iy_i$ as
\begin{align*}
\text{min}\quad &\sum_{i=1}^{k}q_iy_i \\
\text{subject to}\quad & \sum_{i=1}^{k}f(q_i) \leq K,
\\& \sum_{i=1}^{k} q_i =1,
\\& q_i\geq 0 \quad \forall i\in\{1,\ldots,k\}.
\end{align*}
The Lagrangian of this problem is 
\[L(y,\lambda,\beta)=\sum_{i=1}^{k} q_iy_i-\lambda(K-\sum_{i=1}^{k} f (q_i))+\beta(1-\sum_{i=1}^{k} q_i)\]
for $\lambda\geq 0 $ and $\beta \in \R$.
Since Slater's condition holds, using strong duality, we obtain
\begin{align*}
    U(X)&=\inf_{q\geq 0}\sup_{\lambda \geq 0, \beta \in \R} L(y,\lambda,\beta)=\sup_{\lambda \geq 0, \beta \in \R}\inf_{q\geq 0}\ L(y,\lambda,\beta)
\\&= \sup_{\lambda \geq 0, \beta \in \R}\of{-\lambda K+\beta +\inf_{q\geq 0}\sum_{i=1}^{k} \of{(y_i-\beta)q_i + \lambda  f(q_i)}}
\\&=\sup_{\lambda \geq 0, \beta \in \R} \of{-\lambda K+\beta +\sum_{i=1}^{k} \inf_{q_i\geq 0}\of{(y_i-\beta)q_i+\lambda f(q_i)}}
\\&=\sup_{\lambda \geq 0, \beta \in \R} \of{-\lambda K +\beta -\sum_{i=1}^{k} \sup_{q_i \geq 0}\of{(\beta-y_i)q_i-\lambda f(q_i)}}
\\&=\sup_{\beta \in \R}\of{\beta-\sup_{q_i\geq 0}\sum_{i=1}^{k}(\beta-y_i)q_i}\vee \sup_{\lambda > 0, \beta \in \R}\of{-\lambda K +\beta -\lambda \sum_{i=1}^{k} f^\ast\of{\frac{\beta-y_i}{\lambda}}},
\end{align*}
where $\vee$ denotes the maximum. 

First, let us prove the following:
\begin{equation}\label{eq: lambda=0}
  \sup_{\beta \in \R}\of{\beta-\sup_{q_i\geq 0}\sum_{i=1}^{k}(\beta-y_i)q_i}=y_1.  
\end{equation}
Note that for $\beta>y_1$, we can choose $q_2=\ldots=q_{k}=0$ and $q_1 \to \infty$, which makes $\sup_{q_i\geq 0}\sum_{i=1}^{k}(\beta-y_i)q_i=\infty$ and $\beta-\sup_{q_i\geq 0}\sum_{i=1}^{k}(\beta-y_i)q_i=-\infty$. 
Therefore, to find the maximum, we need to look at the case $\beta\leq y_1$. 
Now, $\sup_{q_i\geq 0}\sum_{i=1}^{k}(\beta-y_i)q_i=0$ since $\beta-y_i\leq 0$ for every $i=1,2,\ldots,k$. 
Hence, we obtain \eqref{eq: lambda=0}.

Second, we will show that
\begin{equation}\label{eq: lambda>0}
   \sup_{\lambda > 0, \beta \in \R}\of{-\lambda K +\beta -\lambda \sum_{i=1}^{k} f^\ast\of{\frac{\beta-y_i}{\lambda}}}\geq y_1. 
\end{equation}
Note that Fenchel-Moreau theorem gives that $f(1)=\sup_{x\in \R}(x-f^\ast(x))$.
Since $f(1)>K$, there exists $t\in \R$ and $\epsilon>0$ such that $1\geq t-f^\ast(t)>K+\epsilon$. 
Similarly, the fact that $f(0)=0$ and the Fenchel-Moreau theorem yield that $\inf_{x\in\R}f^\ast(x)=0$. 
Therefore, there exists $z\in \R$ such that $f^\ast(z)<\epsilon/k$. 
If $z\geq t$, we can find $\tilde{z}<t$ such that $f^\ast(\tilde{z})<\epsilon/k$ since $f^\ast$ is a non-decreasing function. 
Therefore, suppose that $z<t$ without loss of generality. 
Now, take $\tilde{\beta}\in\R$ and $\tilde{\lambda}>0$ as the solution of the equations $(\tilde{\beta}-y_1)/\tilde{\lambda}=t$ and $(\tilde{\beta}-y_2)/\tilde{\lambda} = z$. 
Note that $\tilde{\lambda}=(y_2-y_1)/(t-z)>0$. 
The monotonicity of $f^\ast$ implies that 
\[f^\ast\of{\frac{\tilde{\beta}-y_{k}}{\tilde{\lambda}}}\leq \ldots f^\ast\of{\frac{\tilde{\beta}-y_3}{\tilde{\lambda}}}\leq f^\ast\of{\frac{\tilde{\beta}-y_2}{\tilde{\lambda}}}<\frac{\epsilon}{k}. \]
Due to the choice of $\tilde{\lambda}$ and $\tilde{\beta}$, we have
\begin{align*}
-\sum_{i=1}^{k} f^\ast\of{\frac{\tilde{\beta}-y_i}{\tilde{\lambda}}}>K+\epsilon-\frac{\tilde{\beta}-y_1}{\tilde{\lambda}}-\sum_{i=2}^{k} \frac{\epsilon}{k}>K-\frac{\tilde{\beta}-y_1}{\tilde{\lambda}},
\end{align*}
i.e.,
\begin{align*}
    -\tilde{\lambda}K+\tilde{\beta}-\tilde{\lambda}\sum_{i=1}^{k} f^\ast\of{\frac{\tilde{\beta}-y_i}{\tilde{\lambda}}}>y_1,
\end{align*}
which implies \eqref{eq: lambda>0}. 
Therefore, we can write
\begin{equation*}\label{eq: Mp with f(1)>k}
    U(X)=\sup_{\lambda > 0, \beta \in \R}\of{-\lambda K +\beta -\lambda \sum_{i=1}^{k} f^\ast\of{\frac{\beta-y_i}{\lambda}}}.
\end{equation*}

For any $\lambda>0$ and $\beta \in \R$, Lemma~\ref{lem: third-derivative} yields that
\begin{align*}
  &f^\ast\of{\frac{\beta-u_1-\bar{u}}{\lambda}}+f^\ast \of{\frac{\beta-u_2+2\bar{u}}{\lambda}}+f^\ast\of{\frac{\beta-u_3-\bar{u}}{\lambda}} 
  \\& \leq f^\ast\of{\frac{\beta-u_1+\bar{u}}{\lambda}}+f^\ast \of{\frac{\beta-u_2-2\bar{u}}{\lambda}}+f^\ast\of{\frac{\beta-u_3+\bar{u}}{\lambda}},  
\end{align*}
which gives
\begin{align*}
    &-\lambda K +\beta 
    \\&-\lambda\of{f^\ast\of{\frac{\beta-u_1+\bar{u}}{\lambda}}+f^\ast \of{\frac{\beta-u_2-2\bar{u}}{\lambda}}+f^\ast\of{\frac{\beta-u_3+\bar{u}}{\lambda}}+\sum_{i\neq 1,2,3}f^\ast\of{\frac{\beta-u_i}{\lambda}}}
    \\& \leq  -\lambda K +\beta 
    \\&-\lambda\of{f^\ast\of{\frac{\beta-u_1-\bar{u}}{\lambda}}+f^\ast \of{\frac{\beta-u_2+2\bar{u}}{\lambda}}+f^\ast\of{\frac{\beta-u_3-\bar{u}}{\lambda}}+\sum_{i\neq 1,2,3}f^\ast\of{\frac{\beta-u_i}{\lambda}}}.
\end{align*}
Taking the supremum over $\lambda>0$ and $\beta\in \R$, we obtain $U\of{X_A^{(3)}}\leq U\of{X_B^{(3)}}$. 
Hence, the DM is ambiguity prudent.
\end{proof}

\vskip 0.2 cm

\begin{proof}[Proof of Theorem~\ref{thm: aversion for epsilon}]
As is well-known, 
the $\epsilon$-contamination model can be written as the NCEU model with $a=\epsilon$ and $b=\epsilon(2\alpha-1)$.
Therefore, Theorem~\ref{thm:ambavr-NCEU} implies that this model is ambiguity averse if and only if $a=b$, which is equivalent to $\alpha = 1$.
\end{proof}

\vskip 0.2 cm

\begin{proof}[Proof of Theorem~\ref{thm: prudence for epsilon}]
The result is a direct consequence of Theorem~\ref{thm:ambprud-NCEU}, since this model can be written as the NCEU model with $a=\epsilon$ and $b=\epsilon(2\alpha-1)$.
\end{proof}

\subsection{Proofs of Section~\ref{sec:SA}}

\begin{proof} [Proof of Lemma~\ref{lem:IndifferenceforSA}]
Consider $X_I^{(2)}(\omega_1,\omega_2)$ and $X_I^{(2)}(\omega_2,\omega_1).$ 
Take $P_1, P_2 \in M$ such that $P_1(\omega_1)=P_2(\omega_2)=p_1$ and $P_2(\omega_1)=P_1(\omega_2)=p_2$ and they are identical except $\omega_1,\omega_2$. 
Note that $\E_{P_1}[u(X_I^{(2)}(\omega_1,\omega_2))]=\E_{P_2}[u(X_I^{(2)}(\omega_2,\omega_1))]$. 
Due to the symmetry of $\mu$, we have \[\mu(.,p_1,p_2,.)\phi(\E_{P_1}[u(X_I^{(2)}(\omega_1,\omega_2))])=\mu(.,p_2,p_1,.)\phi(\E_{P_2}[u(X_I^{(2)}(\omega_2,\omega_1))]).\] 
Note that for every $P_1\in M$, we can find a corresponding $P_2$ due to Assumption~\ref{Assum: SA}. 
Therefore, we have
    \[\E_{\mu}[\phi(\E_P[u(X_I^{(2)}(\omega_1,\omega_2))])]=\E_{\mu}[\phi(\E_P[u(X_I^{(2)}(\omega_2,\omega_1))])].\]

\end{proof}

\vskip 0.2 cm

\begin{proof}[Proof of Lemma~\ref{lem: phi symm}]
Set $\tilde{\mu}_1:=\mu$ and $\tilde{\mu}_2(.,p_1,p_2,.)=\mu(.,p_2,p_1,.)$. Furthermore, identify acts $X$ and probability measures $P$ with vectors of possible outcomes $u\in\mathbb{R}^k$ (or $\mathbb{R}^\infty$) and $p\in\mathbb{R}^k $ (or $\mathbb{R}^\infty$). 
Using vector product notation we can then write $\mathbb{E}_P[u(X)]=up$, and if $P$ is random (i.e., if $P$ is random variable with distribution $\mu$) we can write $\mathbb{E}_P[u(X)]=uP$. 

Now, the DM being indifferent between permutations of $\omega_1,\omega_2$ and making decisions according to SA model entails that \eqref{eq: eq1} holds. 
Applying then Lemma~\ref{lemma1-add} yields that indeed $\mu(.,p_1,p_2,.)=\mu(.,p_2,p_1,.)$.
\end{proof}	

\vskip 0.2 cm

\begin{proof}[Proof of Lemma~\ref{lemma1-add}]
Suppose by contradiction that $\varphi^{(k)}=0$ for a $k\in \mathbb{N}$ with $\varphi^{(k)}$ being the $k$-th derivative of function $\varphi$. 
Then also $\varphi^{(k+j)}=0$  for $j=1,2,\ldots$ so that $\varphi$ would be a polynomial. 
As $\varphi$ is strictly increasing, the greatest exponent of this polynomial must be uneven, and, since $\varphi$ is not linear, must be greater than $1$. 
However, this is a contradiction to $\varphi$ being concave. 
Hence, no derivative of $\varphi$ can be identical zero. 
In particular, for every $k\in \mathbb{N}$ there exists $ a_k \in \mathbb{R}$ such that $\varphi^{(k)}(a_k)\neq 0$.

Next, define, for $t\in \mathbb{R}$,
\begin{align*}
f_{1,k}(t):&=\mathbb{E}[\varphi(txP_1+a_k)]=\mathbb{E}[\varphi(t(x+\frac{a_ke}{t})P_1)]=\mathbb{E}[\varphi(t(x+\frac{a_k e}{t})P_2)]\\&=\mathbb{E}[\varphi(txP_2+a_k)]=:f_{2,k}(t),
\end{align*} 
with $e=(1,\ldots,1)$, where the second and fourth equations hold as $P_i$ sum up to $1$. 
The third equation holds by \eqref{eq: eq1}. 
Note that for $t=0$ all equations hold setting $\frac{0}{0}=1$. 
Taking the $k$-th derivative of $f_{1,k}$ and $f_{2,k}$, and using that $P_1$ and $P_2$ are bounded (by one), we then have that
\begin{align*}
\mathbb{E}[(xP_1)^k\varphi^{(k)}(a_k)]=f^{(k)}_{1,k}(0)=f^{(k)}_{2,k}(0)=\mathbb{E}[(xP_2)^k\varphi^{(k)}(a_k)].
\end{align*}
Thus, $\mathbb{E}[(xP_1)^k]=\mathbb{E}[(xP_2)^k]$ for all $k\in\mathbb{N}$. 
As $P_1$ and $P_2$ are both uniformly bounded, the power series representation of the exponential function yields that the moment generating functions of $xP_1$ and $xP_2$ must agree. 
Hence, $ xP_1\stackrel{d}{=} xP_2 $ for all $x$, which implies that $P_1\stackrel{d}{=} P_2$, and in particular $\tilde{\mu}_1=\tilde{\mu}_2$.
\end{proof}

\vskip 0.2 cm

\begin{proof}[Proof of Lemma~\ref{lem: SOEU sym}]
    Note that 
    \[U\of{X_I^{(2)}(\omega_1,\omega_2)}-U\of{X_I^{(2)}(\omega_2,\omega_1)}=(P(\omega_1)-P(\omega_2))(\phi(u_1)-\phi(u_2))\]
    needs to be $0$ due to Assumption~\ref{ass:sym}. 
    This implies $P(\omega_1)=P(\omega_2)$, and hence the uniformity.
\end{proof}

\vskip 0.2 cm 

\begin{proof}[Proof of Theorem~\ref{thm: SOEU aversion}]
We calculate the following, using Lemma~\ref{lem: SOEU sym}:
\begin{equation*}
        U\of{X_B^{(2)}}-U\of{X_A^{(2)}}=\frac{1}{k}(\phi(u_1+\bar{u})+\phi(u_2-\bar{u}))-\frac{1}{k}(\phi(u_1-\bar{u})+\phi(u_2+\bar{u})).
\end{equation*}
Therefore, the DM is ambiguity averse if and only if 
\[\phi(u_1+\bar{u})+\phi(u_2-\bar{u})\geq \phi(u_1-\bar{u})+\phi(u_2+\bar{u}),\]
for all $u_1< u_2$ and $\bar{u}>0$,
which is equivalent to the concavity of $\phi$. 
\end{proof}

\vskip 0.2 cm

\begin{proof}[Proof of Theorem~\ref{thm: SOEU prudence}]
Using Lemma~\ref{lem: SOEU sym}, we have
    \begin{align*}
        U\of{X_B^{(3)}}-U\of{X_A^{(3)}}&=\frac{1}{k}(\phi(u_1+\bar{u})+\phi(u_2-2\bar{u})+\phi(u_3+\bar{u}))
        \\&-\frac{1}{k}(\phi(u_1-\bar{u})+\phi(u_2+2\bar{u})+\phi(u_3-\bar{u})).     
    \end{align*}
Therefore, the SOEU DM is ambiguity prudent if and only if 
\[\phi(u_1+\bar{u})+\phi(u_2-2\bar{u})+\phi(u_3+\bar{u})\geq \phi(u_1+\bar{u})+\phi(u_2-2\bar{u})+\phi(u_3+\bar{u}),\]
for any $u_1< u_3$, $u_2=\frac{u_1+u_3}{2}$ and $\bar{u}>0$. 
Then, Lemma~\ref{lem: third-derivative} implies that this inequality is true if and only if $\phi'''\geq 0$, hence the result follows.
\end{proof}

\vskip 0.2 cm 

\begin{proof}[Proof of Theorem~\ref{thm: SA aversion}]
Suppose that $\phi$ is concave. 
Take $P_1, P_2 \in M$ such that $P_1(\omega_1)=P_2(\omega_2)=p_1$ and $P_2(\omega_1)=P_1(\omega_2)=p_2$ and they are identical except $\omega_1,\omega_2$. 
Without loss of generality assume that $p_1\leq p_2$. 
Then, as $u_2-u_1>\bar{u}$, we have
    \begin{align*}
      p_2u_1+p_1u_2+\bar{u}(p_1-p_2)&\leq p_2u_1+p_1u_2 +\bar{u}(p_2-p_1)
      \\&\leq p_1u_1+p_2u_2\leq p_1u_1+p_2u_2+\bar{u}(p_2-p_1).  
    \end{align*}
Using the concavity of the function $\phi$, we obtain
\begin{align*}
        &\phi(.+p_2u_1+p_1u_2+\bar{u}(p_2-p_1)+.)+\phi(.+p_1u_1+p_2u_2+\bar{u}(p_1-p_2)+.)
        \\&\geq \phi(.+p_2u_1+p_1u_2+\bar{u}(p_1-p_2)+.)+\phi(.+p_1u_1+p_2u_2+\bar{u}(p_2-p_1)+.).
\end{align*}
We know that $\mu(P_1)=\mu(P_2)$ due to the symmetry. 
Therefore, this inequality is equivalent to the following:
{\small \[
        \mu(P_2)\phi(\E_{ P_2}[u(X_B^{(2)})])+\mu(P_1)\phi(\E_{ P_1}[u(X_B^{(2)})])\geq \mu(P_2)\phi(\E_{ P_2}[u(X_A^{(2)})])+\mu(P_1)\phi(\E_{P_1}[u(X_A^{(2)})]).
\]
}
Since we can find a corresponding $P_2$ for any $P_1 \in M$, we have
\[\mathbb{E}_{\mu}\sqb{\phi\of{\E_{P}[u(X_B^{(2)})]}}\geq \mathbb{E}_{\mu}\sqb{\phi\of{\E_{P}[u(X_A^{(2)})]}},\]
i.e., ambiguity aversion.

For the other side, suppose that $\phi$ is not concave. 
Then, one can find $0\leq p_1\leq p_2 \leq 1$, $p_1+p_2=1$, $0<u_1 < u_2$ and $\bar{u}>0$ such that
\begin{align*}
         &\phi(p_1u_1+p_2u_2+\bar{u}(p_2-p_1)+\phi(p_2u_1+p_1u_2+\bar{u}(p_1-p_2))
         \\&> \phi(p_1u_1+p_2u_2+\bar{u}(p_1-p_2)+\phi(p_2u_1+p_1u_2+\bar{u}(p_2-p_1)). 
\end{align*}
Consider the following probability measures: $P_1(\omega_1)=p_1$, $P_1(\omega_2)=p_2$ and $P_2(\omega_1)=p_2$, $P_2(\omega_2)=p_1$. 
Take the support of $\mu$ as $\{P_1,P_2\}$. 
Then, $\mu(P_1)=\mu(P_2)=1/2$ follows from the symmetry of $\mu$. 
Since the DM prefers $X_B^{(2)}$ to $X_A^{(2)}$, we obtain
\begin{align*}
         &\mu(P_1)\phi(p_1u_1+p_2u_2+\bar{u}(p_1-p_2)+\mu(P_2)\phi(p_1u_1+p_2u_2+\bar{u}(p_1-p_2)
         \\&\geq \mu(P_1)\phi(p_1u_1+p_2u_2+\bar{u}(p_2-p_1))+\mu(P_2)\phi(p_2u_1+p_1u_2+\bar{u}(p_1-p_2)),
\end{align*}
which contradicts the previous inequality. 
Hence, $\phi$ is concave.
\end{proof}

\vskip 0.2 cm

\begin{proof}[Proof of Theorem~\ref{thm: SA prudence}]
Suppose that $\phi'''\geq 0$ and take $P=P_{(1,2,3)}\in M$. 
Denote $P_{(1,2,3)}(\omega_1)=p_1$, $P_{(1,2,3)}(\omega_2)=p_2,$ $ P_{(1,2,3)}(\omega_3)=p_3$. 
Without loss of generality, assume that $p_1\geq p_2 \geq p_3$. 
Furthermore, define the probability measure $P_{(i,j,k)}$ for any permutation $(i,j,k)$ of the set $\{1,2,3\}$, that satisfy $P_{(i,j,k)}(\omega_1)=p_i$, $P_{(i,j,k)}(\omega_2)=p_j$, $P_{(i,j,k)}(\omega_3)=p_k$ and are identical in the rest. Consider $X_I^{(3)}$, $X_A^{(3)}$ and $X_B^{(3)}$ in \eqref{eq:X_I^3}, \eqref{eq:X_A^3} and \eqref{eq:X_B^3}. 
Let us introduce the following terms with $u_3=u_2+\bar{u}$ and $u_1=u_2-\bar{u}$:
\begin{align*}
    T_1&=.+p_3u_3+p_2u_2+p_1u_1,\\
    T_2&=.+p_2u_3+p_3u_2+p_1u_1,\\
    T_3&=.+p_3u_3+p_1u_2+p_2u_1,\\
    T_4&=.+p_2u_3+p_1u_2+p_3u_1,\\
    T_5&=.+p_1u_3+p_3u_2+p_2u_1,\\
    T_6&=.+p_1u_3+p_2u_2+p_3u_1. 
\end{align*}
Note that 
\begin{align*}
    T_2-T_1=T_6-T_5&=(p_2-p_3)(u_3-u_1)/2 \geq 0,\\
    T_3-T_2=T_5-T_4&=(p_1+p_3-2p_2)(u_3-u_1)/2 \geq 0,\\
    T_4-T_3&=(p_2-p_3)(u_3-u_1)\geq 0.\\
\end{align*}
Furthermore, denote $c=\bar{u}(p_1+p_3-2p_2)$, and $a=\bar{u}(p_2+p_1-2p_3)$. 
Note that $a,c\geq 0$ due to the definition of $M$. 
For $X_A^{(3)}$ and $X_B^{(3)}$, we can write the following:
\begin{align*}
&\E_{P_{(1,2,3)}}[u(X_A^{(3)})]=T_1-c && \E_{P_{(1,2,3)}}[u(X_B^{(3)})]=T_1+c \\
&\E_{P_{(1,3,2)}}[u(X_A^{(3)})]=T_2-a && \E_{P_{(1,3,2)}}[u(X_B^{(3)})]=T_2+a \\
&\E_{P_{(2,1,3)}}[u(X_A^{(3)})]=T_3+a+c && \E_{P_{(2,1,3)}}[u(X_B^{(3)})]=T_3-a-c \\
&\E_{P_{(3,1,2)}}[u(X_A^{(3)})]=T_4+a+c && \E_{P_{(3,1,2)}}[u(X_B^{(3)})]=T_4-a-c \\
&\E_{P_{(2,3,1)}}[u(X_A^{(3)})]=T_5-a && \E_{P_{(2,3,1)}}[u(X_B^{(3)})]=T_5+a \\
&\E_{P_{(3,2,1)}}[u(X_A^{(3)})]=T_6-c && \E_{P_{(3,2,1)}}[u(X_B^{(3)})]=T_6+c. \\
\end{align*}

Consider the following two acts:
\begin{align*}
&T_A=\sqb{T_1-c,\frac{1}{6};T_2-a,\frac{1}{6};T_3+a+c,\frac{1}{6};T_4+a+c,\frac{1}{6};T_5-a,\frac{1}{6};T_6-c,\frac{1}{6}}
\\&T_B=\sqb{T_1+c,\frac{1}{6}; T_2+a,\frac{1}{6};T_3-a-c,\frac{1}{6};T_4-a-c,\frac{1}{6};T_5+a,\frac{1}{6};T_6+c, \frac{1}{6}} 
\end{align*} 
for the Expected Utility DM with utility function $\phi$.
Note that $\E[T_{A}]=\E[T_{B}]$ and $\E[T_{A}^2]=\E[T_{B}^2]$. $\E[T_A]=\E[T_B]$ implies that $\int_0^{T_6+c}F_{T_A}(z)\diff z = \int_0^{T_6+c}F_{T_B}(z)\diff z$. 
Similarly, the equality of second moments gives that $\int_0^{T_6+c}\int_0^yF_{T_A}(z)\diff z \diff y = \int_0^{T_6+c}\int_0^yF_{T_B}(z)\diff z \diff y$ due to \citet[Proposition 1]{DE10}.
We can calculate the following:
\begin{align}\label{eq: Second order diff}
        \int_0^y (F_{T_A}(z)-F_{T_B}(z))\diff z = \begin{cases}
            0 &\text{if } y<T_1-c,\\
            \frac{y-(T_1-c)}{6} &\text{if } T_1-c\leq y <T_1+c,\\
            \frac{2c}{6} & \text{if } T_1+c\leq y < T_2-a,\\
            \frac{2c}{6}+\frac{y-(T_2-a)}{6} &\text{if } T_2-a\leq y <T_2+a,\\
            \frac{2c+2a}{6}&\text{if } T_2+a\leq T_3-a-c,\\
            \frac{2c+2a}{6}-\frac{y-(T_3-a-c)}{6}&\text{if } T_3-a-c\leq y <T_3+a+c,\\
            0 &\text{if } T_3+a+c \leq y<T_4-a-c,\\
            -\frac{y-(T_4-a-c)}{6} &\text{if } T_4-a-c\leq y<T_4+a+c,\\
            -\frac{2a+2c}{6} &\text{if } T_4+a+c \leq y < T_5-a,\\
            \frac{y-(T_5-a)}{6}-\frac{2a+2c}{6}&\text{if }T_5-a\leq y<T_5+a,\\
            -\frac{2c}{6} &\text{if } T_5+a\leq y <T_6-c,\\
            \frac{y-(T_6-c)}{6}-\frac{2c}{6} &\text{if } T_6-c\leq y<T_6+c,\\
            0 &\text{if } T_6+c\leq y.
        \end{cases}
\end{align}
Furthermore, we can see from \eqref{eq: Second order diff} that
\[\int_0^x\int_0^y\of{F_{T_{A}}(z)-F_{T_{B}}(z)}\diff z \diff y \geq 0\]
for all $x$ since the integrated term $\int_0^y\of{F_{T_{A}}(z)-F_{T_{B}}(z)}\diff z$ is nonnegative before $T_4-a-c$ and nonpositive afterwards, and the integral 
\[\int_0^{T_6+c}\int_0^x(F_{T_A}(y)-F_{T_B}(y))\diff y \diff x\] equals $0$ at the end value. 
Therefore, Theorems~1 and~2 in \cite{MGT80} imply that $\E[\phi(T_{B})]\geq \E[\phi(T_{A})]$ since $\phi'''\geq 0$. 
This inequality can be rewritten as:
\begin{align*}
     &\phi(T_1+c)+\phi(T_2+a)+\phi(T_3-a-c)+\phi(T_4-a-c)+\phi(T_5+a)+\phi(T_6+c)
    \\&\geq \phi(T_1-c)+\phi(T_2-a)+\phi(T_3+a+c)+\phi(T_4+a+c)+\phi(T_5-a)+\phi(T_6-c),
\end{align*}
which is
\begin{align*}
       &\phi\of{\E_{P_{(1,2,3)}}[u(X_B^{(3)})]}
       +\phi\of{\E_{P_{(1,3,2)}}[u(X_B^{(3)})]}
       +\phi\of{\E_{P_{(2,1,3)}}[u(X_B^{(3)})]}
       \\+&\phi\of{\E_{P_{(3,1,2)}}[u(X_B^{(3)})]}
       +\phi\of{\E_{P_{(2,3,1)}}[u(X_B^{(3)})]}
       +\phi\of{\E_{P_{(3,2,1)}}[u(X_B^{(3)})]}
       \\\geq &\phi\of{\E_{P_{(1,2,3)}}[u(X_A^{(3)})]}
       +\phi\of{\E_{P_{(1,3,2)}}[u(X_A^{(3)})]}
       +\phi\of{\E_{P_{(2,1,3)}}[u(X_A^{(3)})]}
       \\+&\phi\of{\E_{P_{(3,1,2)}}[u(X_A^{(3)})]}
       +\phi\of{\E_{P_{(2,3,1)}}[u(X_A^{(3)})]}
       +\phi\of{\E_{P_{(3,2,1)}}[u(X_A^{(3)})]}.
\end{align*}
Due to the symmetry of $\mu$,
$\mu(P_{(1,2,3)})=\mu(P_{(1,3,2)})=\mu(P_{(2,3,1)})=\mu(P_{(2,1,3)})=\mu(P_{(3,2,1)})=\mu(P_{(3,1,2)})$. 
Therefore, we obtain

\begin{align*}
       &\mu(P_{(1,2,3)})\phi\of{\E_{P_{(1,2,3)}}[u(X_B^{(3)})]}
       +\mu(P_{(1,3,2)})\phi\of{\E_{P_{(1,3,2)}}[u(X_B^{(3)})]}
       \\+&\mu(P_{(2,1,3)})\phi\of{\E_{P_{(2,1,3)}}[u(X_B^{(3)})]}
       +\mu(P_{(3,1,2)})\phi\of{\E_{P_{(3,1,2)}}[u(X_B^{(3)})]}
       \\+&\mu(P_{(2,3,1)})\phi\of{\E_{P_{(2,3,1)}}[u(X_B^{(3)})]}
       +\mu(P_{(3,2,1)})\phi\of{\E_{P_{(3,2,1)}}[u(X_B^{(3)})]}
       \\\geq &\mu(P_{(1,2,3)})\phi\of{\E_{P_{(1,2,3)}}[u(X_A^{(3)})]}
       +\mu(P_{(1,3,2)})\phi\of{\E_{P_{(1,3,2)}}[u(X_A^{(3)})]}
       \\+&\mu(P_{(2,1,3)})\phi\of{\E_{P_{(2,1,3)}}[u(X_A^{(3)})]}
       +\mu(P_{(3,1,2)})\phi\of{\E_{P_{(3,1,2)}}[u(X_A^{(3)})]}
       \\+&\mu(P_{(2,3,1)})\phi\of{\E_{P_{(2,3,1)}}[u(X_A^{(3)})]}
       +\mu(P_{(3,2,1)})\phi\of{\E_{P_{(3,2,1)}}[u(X_A^{(3)})]}.
\end{align*}
Since this is valid for any $P \in M$ and we can find the relevant permutations for any $P\in M$, this leads to
\[\E_{\mu}[\phi(\E_P[u(X_B^{(3)})])]\geq \E_{\mu}[\phi(\E_P[u(X_A^{(3)})])].\] 

To prove the converse, take the support of $\mu$ as
\[
\cb{P_{(1,2,3)},P_{(1,3,2)},P_{(2,3,1)},P_{(2,1,3)},P_{(3,2,1)},P_{(3,1,2)}}.
\]
Then, the symmetry of $\mu$ implies that \[\mu(P_{(1,2,3)})=\mu(P_{(1,3,2)})=\mu(P_{(2,3,1)})=\mu(P_{(2,1,3)})=\mu(P_{(3,2,1)})=\mu(P_{(3,1,2)})=\frac{1}{6}.\]
Therefore, ambiguity prudence, i.e., $\E_{\mu}[\phi(\E_P[u(X_B^{(3)})])]\geq \E_{\mu}[\phi(\E_P[u(X_A^{(3)})])]$, implies that \[\E[\phi(T_B)]\geq \E[\phi(T_A)]\] for any choice of $X_A^{(3)}$ and $X_B^{(3)}$ of the form \eqref{eq:X_A^3} and \eqref{eq:X_B^3}. 
Note that $P_{(1,2,3)}$ is also arbitrarily chosen. 
Using integration by parts, we obtain
\begin{align*}
&\E[\phi(T_B)]-\E[\phi(T_A)]=\int_{0}^{T_6+c}\phi(x)\diff F_{T_B}(x)-\int_{0}^{T_6+c}\phi(x)\diff F_{T_A}(x) 
\\&=\of{\phi(T_6+c)F_{T_B}(T_6+c)-\int_0^{T_6+c}\phi'(x)F_{T_B}(x)\diff z}
\\&-\of{\phi(T_6+c)F_{T_A}(T_6+c)-\int_0^{T_6+c}\phi'(x)F_{T_A}(x)\diff z}
\\&=\int_0^{T_6+c}\phi'(x)(F_{T_A}(x)-F_{T_B}(x))\diff x
\\&=\phi'(T_6+c)\int_0^{T_6+c}(F_{T_A}(x)-F_{T_B}(x))\diff x - \int_0^{T_6+c}\phi''(x)\int_0^x (F_{T_A}(y)-F_{T_B}(y))\diff y \diff x
\\&=- \int_0^{T_6+c}\phi''(x)\int_0^x (F_{T_A}(y)-F_{T_B}(y))\diff y \diff x
\\&=-\phi''(T_6+c)\int_0^{T_6+c}\int_0^x(F_{T_A}(y)-F_{T_B}(y))\diff y \diff x
\\&+ \int_0^{T_6+c}\phi'''(x)\int_0^x\int_0^y (F_{T_A}(z)-F_{T_B}(z))\diff z\diff y \diff x
\\&= \int_{0}^{\infty}\phi'''(x)\int_0^x\int_0^y\of{F_{T_{A}}(z)-F_{T_{B}}(z)}\diff z \diff y \diff x,
\end{align*}
where we used $1=F_{T_B}(T_6+c)=F_{T_A}(T_6+c)$ in the fourth line, $\E[T_A] = \E[T_B]$ (i.e., $\int_0^{T_6+c}(F_{T_A}(x)-F_{T_B}(x))\diff x=0$) in the sixth line, and $\E[T_B^2]=\E[T_A^2]$ (i.e., $\int_0^{T_6+c}\int_0^x(F_{T_A}(y)-F_{T_B}(y))\diff y \diff x =0$) in the ninth line. 
Since this equality is valid for any choice of $X_B$ and $X_A$ in the form of \eqref{eq:X_A^3} and \eqref{eq:X_B^3}, $\int_0^x\int_0^y\of{F_{T_{A}}(z)-F_{T_{B}}(z)}\diff z \diff y$ can take any positive value. 
Therefore, negative values of $\phi'''$ would make a negative value for $\E[\phi(T_B)]-\E[\phi(T_A)]$ feasible, which is a contradiction. 
Hence, $\phi'''\geq 0$.
\end{proof}

\vskip 0.2 cm

\begin{proof}[Proof of Theorem~\ref{thm: SA prudence 2}]
When $M \subseteq \cb{(.,p_1,p_2,p_3,.)|p_{(2)}\leq \frac{p_{(1)}+ p_{(3)}}{2}}$, Theorem~\ref{thm: SA prudence} gives the ambiguity prudence. 
For the other side, suppose that \[M \not\subseteq \cb{(.,p_1,p_2,p_3,.)|p_{(2)}\leq \frac{p_{(1)}+ p_{(3)}}{2}}.\] 
Then, there exist $p_1\geq p_2 \geq p_3$ such that $p_2> \frac{p_1+p_3}{2}$. 
Choose the support of $\mu$ as this probability measure and its permutations. 
Then, the symmetry of $\mu$ implies that $\mu(.,p_i,p_j,p_k,.)=1/6$ for all permutations $(i,j,k)$ of the set $\{1,2,3\}$. 
Using the same $T_i$'s from the proof of Theorem~\ref{thm: SA prudence} and since $(.,p_i,p_j,p_k,.) \notin \cb{(.,p_1,p_2,p_3,.)|p_{(2)}\leq \frac{p_{(1)}+ p_{(3)}}{2}}$, we have the following inequalities:  $T_1\leq T_3\leq T_2 \leq T_5 \leq T_4\leq T_6$. 
Denote $-c=\bar{u}(p_1+p_3-2p_2)$, and $b=\bar{u}(2p_1-p_2-p_3)$. 
Note that $b,c\geq 0$. 
Consider the following two acts:
\begin{align*}
         &T_A=[T_1+c,\frac{1}{6};T_3+b,\frac{1}{6};T_2-b-c,\frac{1}{6};T_5-b-c,\frac{1}{6};T_4+b,\frac{1}{6};T_6+c, \frac{1}{6}],
         \\&T_B=[T_1-c,\frac{1}{6}; T_3-b,\frac{1}{6};T_2+b+c,\frac{1}{6};T_5+b+c,\frac{1}{6};T_4-b,\frac{1}{6};T_6-c,\frac{1}{6}].
\end{align*}
Since $\phi'''\geq 0$, $\E[T_A]=\E[T_B]$, $\E[T_A^2]=\E[T_B^2]$ and as before
\[\int_0^x\int_0^y\of{F_{T_{A}}(z)-F_{T_{B}}(z)}\diff z \diff y \geq 0\]
holds for all $x$, Theorems~1 and~2 in \cite{MGT80} imply that $\E[\phi(T_A)]> \E[\phi(T_B)]$. 
Note that this is equivalent to 
      \[\E_{\mu}[\phi(\E_P[u(X_B^{(3)})])]< \E_{\mu}[\phi(\E_P[u(X_A^{(3)})])],\] 
      which contradicts ambiguity prudence.
\end{proof}

\subsection{Supplement to Section~\ref{sec: Insurance g-div}}

\subsubsection{Non-Affine Utility}\label{sec:nonaffine}
Suppose now that the DM has a non-affine utility function. 
In this case, the noisy loss is assumed to take the values $\ell-\epsilon_1$ or $\ell+\epsilon_2$, $\epsilon_{1},\epsilon_{2}>0$, with unknown probabilities. 
The values of $\epsilon_1$ and $\epsilon_2$ will be specified later. 
The DM then solves: 
\begin{equation}\label{eq: nonlinear g-div example with noise}
    \max_{0\leq s\leq\ell}\min_{Q\in \Delta}\left\{ \E_Q[u(X_\epsilon(s))]+I_g(Q|P)\right\},
\end{equation}
where
\begin{equation*}
X_\epsilon(s):=[w-\ell-\epsilon_2+s-\pi(s),\omega_{(1,\epsilon_2)};w-\ell+\epsilon_1+s-\pi(s),\omega_{(1,-\epsilon_1)}; w-\pi(s),\omega_2;\ldots;w-\pi(s),\omega_k].
\end{equation*}
To solve this problem, we first define the utility units $\bar{w},\bar{\ell},\bar{s},[\bar{\pi}(\bar{s})]$ as solutions to the following equations:
\begin{align*}
-\bar{\ell}+\bar{s}&=u(w-\ell+s-\pi(s))-u(w-\pi(s)),
\\ \bar{\pi}(\bar{s})&=u(\pi(s)),
\\ \bar{w}-\bar{\pi}(\bar{s})&=u(w-\pi(s)).
\end{align*}
Then, for a fixed $\bar{\epsilon}>0$, we choose the noise $(-\epsilon_1,\epsilon_2)$ as solutions to the following equations:
\[
    -\bar{\epsilon}=u(w-\ell-\epsilon_2+s-\pi(s))-\bar{w}+\bar{\ell}-\bar{s}+\bar{\pi}(\bar{s})=u(w-\ell-\epsilon_2+s-\pi(s))-u(w-\ell+s-\pi(s)),
 \]
\[ 
 \hspace{0.27cm} \bar{\epsilon}=u(w-\ell+\epsilon_1+s-\pi(s))-\bar{w}+\bar{\ell}-\bar{s}+\bar{\pi}(\bar{s})=u(w-\ell+\epsilon_1+s-\pi(s))-u(w-\ell+s-\pi(s)).
\]
Provided the function $u$ is continuous and increasing, finding such $\epsilon_1,\epsilon_2>0$ is always possible. 
Then, \eqref{eq: nonlinear g-div example with noise} can be re-written as
\begin{equation*}
\max_{0\leq\bar{s}\leq\bar{\ell}}\min_{Q\in\Delta}\left\{\E_Q[\bar{X}(\bar{s})]+I_g(Q|P)\right\},
\end{equation*}
where
\begin{equation*}
\bar{X}(\bar{s}):=[\bar{w}-\bar{\ell}-\bar{\epsilon}+\bar{s}-\bar{\pi}(\bar{s}),\omega_{(1,\bar{\epsilon})};\bar{w}-\bar{\ell}+\bar{\epsilon}+\bar{s}-\bar{\pi}(\bar{s}),\omega_{(1,-\bar{\epsilon})};\bar{w}-\bar{\pi}(\bar{s}),\omega_2;\ldots;\bar{w}-\bar{\pi}(\bar{s}),\omega_k].
\end{equation*}
This is effectively the same problem as in the affine utility case, with monetary units replaced by utils.

\subsubsection{Insuring an Ambiguous Loss under the SOEU Model}\label{sec:appSOEU}

In this subsection, we consider the same problem as in Section~\ref{sec: Insurance g-div} for an ambiguity averse DM who complies with the SOEU model. 
Recall that the DM pays $\pi(s)$ for a compensation amount $s$.
It is natural to assume that $s-\pi(s)$ is increasing with respect to $s$, since the compensation amount increases faster than the premium. 
Now the DM wants to solve the following problem:
\begin{equation*}
    \max_{0\leq s \leq \ell}\E[\phi(u(X(s)))],
\end{equation*}
where\footnote{Here, one may invoke the same construction as in Section~\ref{sec:nonaffine}.} 
\begin{equation*}
u(X(s)):= [w-\ell+s-\pi(s),\omega_1;w-\pi(s),\omega_2;\ldots;w-\pi(s),\omega_k].
\end{equation*}
Under Assumption~\ref{ass:sym}, using Lemma~\ref{lem: SOEU sym}, we can rewrite the problem as:
\begin{equation*}
    \max_{0\leq s \leq \ell} \frac{1}{k}\phi(w-\ell+s-\pi(s))+\frac{k-1}{k}\phi(w-\pi(s)).
\end{equation*}
Denote the optimal solution of this problem by $s^\ast$. 
It needs to satisfy the first-order condition, that is, 
\begin{equation*}
(1-\pi'(s^\ast))\phi'(w-\ell+s^\ast-\pi(s^\ast))=(k-1)\pi'(s^\ast)\phi'(w-\pi(s^\ast)).
\end{equation*}
Rewriting this equation, we obtain
\begin{equation}\label{eq: SOEU wtihout noise}
 \frac{\phi'(w-\ell+s^\ast-\pi(s^\ast))}{k-1}=\frac{\pi'(s^\ast)}{1-\pi'(s^\ast)}\phi'(w-\pi(s^\ast)).
\end{equation}
Note that Theorem~\ref{thm: SOEU aversion} concluded that $\phi$ is concave when the DM is ambiguity averse. 
Hence, $\phi'$ is a decreasing function. 
Similarly, the convexity of $\pi$ implies the increasingness of $\pi'$. 
Finally, $\pi'\leq 1$ since $s-\pi(s)$ is increasing. 
Combining these observations, it can be seen that the right-hand side, given by 
\[\frac{\pi'(s)}{1-\pi'(s)}\phi'(w-\pi(s)),\]
is an increasing function of $s$.

Now consider the problem with the noise in the loss amount. 
Then, the DM solves the following problem:
\[\max_{0\leq s \leq \ell} \E[\phi(u(X_\epsilon(s)))],\]
where
\[u(X_\epsilon(s)):=[w-\ell-\epsilon+s-\pi(s),\omega_{(1,\epsilon)};w-\ell+\epsilon+s-\pi(s),\omega_{(1,-\epsilon)}; w-\pi(s),\omega_2;\ldots;w-\pi(s),\omega_k].\]
Under Assumption~\ref{ass:sym}, we can rewrite the problem as:
\[\max_{0\leq s \leq \ell}\frac{1}{2k}\phi(w-\ell-\epsilon+s-\pi(s))+\frac{1}{2k}\phi(w-\ell+\epsilon+s-\pi(s))+\frac{k-1}{k}\phi(w-\pi(s)).\]
Denote the optimal solution of this problem by $s^\ast_\epsilon$. 
It needs to satisfy the first-order condition, that is,
\[\frac{1-\pi'(s^\ast_{\epsilon})}{2}(\phi'(w-\ell-\epsilon+s^\ast_{\epsilon}-\pi(s^\ast_{\epsilon}))+\phi'(w-\ell+\epsilon+s^\ast_{\epsilon}-\pi(s^\ast_{\epsilon})))=(k-1)\pi'(s^\ast_{\epsilon})\phi'(w-\pi(s^\ast_{\epsilon})).\]
Rewriting this expression, we obtain
\begin{equation}\label{eq: SOEU with noise}
    \frac{\phi'(w-\ell-\epsilon+s^\ast_{\epsilon}-\pi(s^\ast_{\epsilon}))+\phi'(w-\ell+\epsilon+s^\ast_{\epsilon}-\pi(s^\ast_{\epsilon}))}{2(k-1)}=\frac{\pi'(s^\ast_{\epsilon})}{1-\pi'(s^\ast_{\epsilon})}\phi'(w-\pi(s^\ast_{\epsilon})).
\end{equation}
Now suppose that $s^\ast >s^\ast_\epsilon$, which is equivalent to
\begin{equation}\label{eq: SOEU contradiction}
    \frac{\pi'(s^\ast_{\epsilon})}{1-\pi'(s^\ast_{\epsilon})}\phi'(w-\pi(s^\ast_{\epsilon}))< \frac{\pi'(s^\ast)}{1-\pi'(s^\ast)}\phi'(w-\pi(s^\ast)).
\end{equation}
Using \eqref{eq: SOEU wtihout noise} and \eqref{eq: SOEU with noise}, we have
\begin{align*}
 \frac{\pi'(s^\ast)}{1-\pi'(s^\ast)}\phi'(w-\pi(s^\ast)) &=  \frac{\phi'(w-\ell+s^\ast-\pi(s^\ast))}{k-1}
  \\&\leq \frac{ \phi'(w-\ell-\epsilon+s^\ast-\pi(s^\ast))+ \phi'(w-\ell+\epsilon+s^\ast-\pi(s^\ast))}{2(k-1)}
  \\& \leq  \frac{\phi'(w-\ell-\epsilon+s^\ast_{\epsilon}-\pi(s^\ast_{\epsilon}))+\phi'(w-\ell+\epsilon+s^\ast_{\epsilon}-\pi(s^\ast_{\epsilon}))}{2(k-1)} 
  \\&= \frac{\pi'(s^\ast_{\epsilon})}{1-\pi'(s^\ast_{\epsilon})}\phi'(w-\pi(s^\ast_{\epsilon})),
\end{align*}
where we have used the convexity of $\phi'$ in the second line and the decreasingness of $\phi'$ in the third line. 
Hence, we obtain  
\[\frac{\pi'(s^\ast_{\epsilon})}{1-\pi'(s^\ast_{\epsilon})}\phi'(w-\pi(s^\ast_{\epsilon})) \geq \frac{\pi'(s^\ast)}{1-\pi'(s^\ast)}\phi'(w-\pi(s^\ast)),\]
which contradicts \eqref{eq: SOEU contradiction}. 
Therefore, $s^\ast_\epsilon \geq s^\ast$ when $\phi'$ is convex. 
We can conclude again that, \textit{ceteris paribus}, the prudent DM is willing to buy more insurance when there is uncertainty with respect to the loss amount.

\vskip 0.2cm

\bibliographystyle{apalike}
\bibliography{ALSreferences}

\end{document}